\documentclass[epj,nopacs]{svjour}

\usepackage{amsmath}
\usepackage{amssymb}
\usepackage[pdftex]{graphicx}
\usepackage[numbers,sort&compress]{natbib}

\usepackage{enumitem}
\setlist[enumerate]{leftmargin=0pt, wide=0.5\parindent, topsep=\parskip, itemsep=\topsep}

\newcommand{\Dmq}{\Delta m^2}
\newcommand{\Eps}{\varepsilon}
\newcommand{\eVq}{\ensuremath{\text{eV}^2}}
\newcommand{\diag}{\mathop{\mathrm{diag}}}
\newcommand{\Nuc}[2]{\ensuremath{^{#1}\text{#2}}}
\renewcommand{\Re}{\mathop{\mathrm{Re}}}

\begin{document}


\title{Solar neutrinos and neutrino physics}

\author{Michele Maltoni\inst{1} \and Alexei Yu.\ Smirnov\inst{2,3}}

\institute{
  Instituto de F\'{\i}sica Te\'orica UAM/CSIC, Calle de
  Nicol\'as Cabrera 13--15, Universidad Aut\'onoma de Madrid,
  Cantoblanco, E-28049 Madrid, Spain
  \and
  Max-Planck Institute for Nuclear Physics, Saupfercheckweg 1, D-69117
  Heidelberg, Germany
  \and
  ICTP, Strada Costiera 11, 34014 Trieste, Italy
}

\date{Review prepared for the European Physical Journal A (EPJA) issue
  on ``The Solar Neutrinos''.}

\abstract{Solar neutrino studies triggered and largely motivated the
  major developments in neutrino physics in the last 50 years. Theory
  of neutrino propagation in different media with matter and fields
  has been elaborated.  It includes oscillations in vacuum and matter,
  resonance flavor conversion and resonance oscillations, spin and
  spin-flavor precession, \textit{etc}.  LMA MSW has been established
  as the true solution of the solar neutrino problem.  Parameters
  $\theta_{12}$ and $\Dmq_{21}$ have been measured; $\theta_{13}$
  extracted from the solar data is in agreement with results from
  reactor experiments.  Solar neutrino studies provide a sensitive way
  to test theory of neutrino oscillations and conversion.
  Characterized by long baseline, huge fluxes and low energies they
  are a powerful set-up to search for new physics beyond the standard
  $3\nu$ paradigm: new neutrino states, sterile neutrinos,
  non-standard neutrino interactions, effects of violation of
  fundamental symmetries, new dynamics of neutrino propagation, probes
  of space and time.  These searches allow us to get stringent, and in
  some cases unique bounds on new physics.  We summarize the results
  on physics of propagation, neutrino properties and physics beyond
  the standard model obtained from studies of solar neutrinos.}

\maketitle


\section{Introduction}

``If the oscillation length is large\dots\ from the point of view of
detection possibilities an ideal object is the Sun.'' This statement
from Pontecorvo's 1967 paper~\cite{Pontecorvo:1967fh} published before
release of the first Homestake experiment results~\cite{Davis:1968cp}
can be considered as the starting point for the solar neutrino studies
of new physics.

Observation of the deficit of signal in the Homestake experiment was
the first indication of existence of oscillations.  This result had
triggered vast experimental~\cite{chapt3} and theoretical developments
in neutrino physics.  On theoretical side, various non-standard
properties of neutrinos have been introduced and new effects in
propagation of neutrinos have been proposed.  These include:
\begin{enumerate}
\item Neutrino spin-precession in the magnetic fields of the Sun due
  to large magnetic moments of neutrinos~\cite{Cisneros:1970nq,
    Okun:1986na}: electromagnetic properties of neutrinos have been
  studied in details.

\item Neutrino decays: Among various possibilities (radiative, $3\nu$
  decay, \textit{etc.}) the decay into light scalar, \textit{e.g.},
  Majoron, is less restricted~\cite{Bahcall:1972my,
    Berezhiani:1987gf}.

\item The MSW effect: The resonance flavor conversion inside the Sun
  required neutrino mass splitting in the range $\Dmq = (10^{-7} -
  10^{-4})~\eVq$ and mixing $\sin^2 2\theta > 10^{-3}$~\cite{w1a, w1b,
    w2, ms1a, ms1b, ms2}.  This was the first correct estimation of
  the neutrino mass and mixing intervals.  With adding more
  information three regions of $\Dmq$ and $\sin^2 2\theta$ have been
  identified: the so called SMA, LMA and LOW solutions.

\item ``Just-so'' solution: vacuum oscillations with nearly maximal
  mixing and oscillation length comparable with distance between the
  Sun and the Earth have been proposed~\cite{Glashow:1987jj}.

\item Oscillations and flavor conversion due to non-standard neutrino
  interactions of massless neutrinos~\cite{w1a, w1b, Roulet:1991sm,
    Guzzo:1991hi}.

\item Resonant spin-flavor precession~\cite{Lim:1987tk,
  Akhmedov:1988uk}, which employs matter effect on neutrino spin
  precession in the magnetic fields.  The effect is similar to the MSW
  conversion.

\item Oscillation and conversion in matter due to violation of the
  equivalence principle~\cite{Gasperini:1988zf}, Lorentz violating
  interactions~\cite{Kostelecky:2003cr}, \textit{etc}.
\end{enumerate}

In turn, these proposals led to detailed elaboration of theory of
neutrino propagation in different media as well as to model-building
which explains non-standard neutrino properties.

Studies of the solar neutrinos and results of KamLAND
experiment~\cite{Eguchi:2002dm, Araki:2004mb, Abe:2008aa} led to
establishing the LMA MSW solution as the solution of the solar
neutrino problem.  Other proposed effects are not realized as the main
explanation of the data.  Still they can be present and show up in
solar neutrinos as sub-leading effects.  Their searches allow us to
get bounds on corresponding neutrino parameters.  Thus, the Sun can be
used as a source of neutrinos for exploration of non-standard neutrino
properties.

In this review we summarize implications of results from the solar
neutrino studies for neutrino physics, the role of solar neutrinos in
establishing the $3\nu$ mixing paradigm, in searches for new physics
beyond the standard model.  The paper is organized as follows. In
Sec.~\ref{sec:msw} physics of the LMA MSW solution of the solar
neutrino problem is described.  We discuss properties of this solution
and dependence of the observables on neutrino parameters.  In
Sec.~\ref{sec:param} determination of the neutrino masses and mixing
using solar neutrinos is described.  We outline status of the solution
and summarize existing open questions.  Sec.~\ref{sec:bsm} is devoted
to possible manifestations of sub-leading effects due to physics
beyond the standard model.  Bounds on parameters of this new physics
are presented.


\section{Propagation and flavor evolution of the solar neutrinos.
  LMA MSW solution}
\label{sec:msw}


\subsection{Evolution. Three phases}
\label{sec:evol}

Evolution of the flavor neutrino states, $\nu_f \equiv (\nu_e,
\nu_\mu, \nu_\tau)^T$, is described by the equation
\begin{equation}
  \label{eq:evolution}
  i \frac{d \nu_f}{dx} = H \nu_f = (H_0 + V) \nu_f \,,
\end{equation}
where $H$ is the total Hamiltonian, $H_0 \approx M^{\dagger}M/2p$ is
the Hamiltonian in vacuum, $M$ is the mass matrix of neutrinos (the
term proportional to the neutrino momentum $p$ is omitted here), and
$V = \diag(V_e, 0, 0)$ is the diagonal matrix of matter potentials
with $V_e = \sqrt{2} G_F n_e$~\cite{w1a, w1b}. Here $G_F$ is the Fermi
constant and $n_e$ is the number density of electrons.

The flavor evolution is described in terms of the instantaneous
eigenstates of the Hamiltonian in matter $\nu_m \equiv (\nu_{1m},
\nu_{2m}, \nu_{3m})^T$.  These eigenstates are related to the flavor
states by the mixing matrix in matter, $U^m$:
\begin{equation}
  \label{eq:mixing}
  \nu_f = U^m \nu_m \,.
\end{equation}
The matrix $U^m$ is determined via diagonalization of the Hamiltonian:
\begin{equation}
  \label{eq:eigenvalues}
  U^{m\dagger} H U^m = H^\text{diag} = \diag(H_{1m}, H_{2m}, H_{3m}) \,,
\end{equation}
where $H_{im}$ are the eigenvalues of the Hamiltonian.  In vacuum
$\nu_{im}$ coincide with the mass eigenstates: $\nu_{im} = \nu_i$, and
$H_{im} \approx m_i^2/2p$.

The physical picture of neutrino propagation and flavor evolution is
the following:
\begin{itemize}
\item Neutrino state produced as $\nu_e$ in the central regions of the
  Sun propagates as the system of eigenstates of the Hamiltonian,
  $\nu_{im}$. Admixtures of the eigenstates are determined by the
  mixing in matter in the production region. The eigenstates propagate
  \emph{independently} of each other and transform into corresponding
  mass eigenstates when arriving at the surface of the Sun: $\nu_{im}
  \to \nu_i$.

\item The mass eigenstates propagate without changes to the surface of
  the Earth. The coherence between these states is lost and
  oscillations are irrelevant.

\item Entering the Earth the mass states $\nu_i$ split (decomposed)
  into the eigenstates in matter of the Earth and oscillate
  propagating inside the Earth to a detector.
\end{itemize}
We will discuss these three phases in the next section.


\subsection{Propagation inside the Sun. Adiabatic flavor conversion}
\label{sec:evol-sun}

The picture of flavor transitions in the Sun is simple.  In the LMA
case the solution of the evolution equation~\eqref{eq:evolution} is
trivial due to large mixing and relatively slow (adiabatic) change of
density on the way of neutrinos.  Namely, in the Sun the condition of
smallness of the density gradient, \textit{i.e.}, the adiabaticity
condition,
\begin{equation}
  \label{eq:adiab}
  d \equiv n_e \left(\frac{d n_e }{dx} \right)^{-1} > l_m \,,
\end{equation}
is satisfied.  Here $l_m = 2 \pi/\Delta_m $ is the oscillation length
in matter and $\Delta_m \equiv H_{im} - H_{jm}$ is the difference of
eigenvalues of the Hamiltonian. According to Eq.~\eqref{eq:adiab}, a
system characterized by the eigenlength $l_m$ has time to adjust
itself to the change of external conditions determined by the scale of
density change, $d$.
Then with high accuracy the solution of Eq.~\eqref{eq:evolution} is
given by the first order adiabatic perturbation theory which we call
the adiabatic solution~\cite{ms2, Bethe:1986ej, Messiah:1986fc}.

The adiabatic solution can be written immediately using the physical
picture outlined in Sec.~\ref{sec:evol}.  The state of electron
neutrino produced in the central regions of the Sun can be decomposed
in terms of the eigenstates in matter as
\begin{equation}
  \label{eq:nue}
  \nu_e = \sum_i U^m_{ei}(n_e^0)~\nu_{im} (n_e^0) \,,
\end{equation}
where $U^m_{ei}(n_e^0)$ are the elements of mixing matrix in the
production point with density $n_e^0$.
The adiabatic evolution means that transitions between the eigenstates
in the course of propagation are negligible and the eigenstates evolve
independently of each other.  So,  the evolution is reduced to (i)
change of $\nu_{jm}$ flavor content and (ii) appearance of the phase
factors
\begin{equation}
  \label{eq:solutionh}
  \nu_{jm}(t_0) \to e^{i \phi_j (t)} \nu_{jm}(t) \,,
\end{equation}
where the phases equal
\begin{equation}
  \phi_j(t) = \int_0^t dz H_{jm}(z) \,.
\end{equation}
The flavor content of the eigenstate in matter changes according to change
of mixing:
\begin{equation}
  \label{eq:flavcont}
  \nu_{jm}(t) = U_{\alpha j}^{m \dagger}(t) \nu_{\alpha} \,,
\end{equation}
and $U_{\alpha j}^m (t) = U_{\alpha j}^m (n(t))$ follows the density
change.  Thus, admixtures of the eigenstates are conserved being fixed
by~\eqref{eq:nue}, but flavors of the eigenstates do change.

At the surface of the Sun we have $\nu_{jm}(t_s) = \nu_j$ and the
neutrino state becomes
\begin{equation}
  \label{eq:nuesurf}
  \nu (t_s)
  = \sum_j U^m_{ej}(n^0) e^{i \phi_j (t_s)} \nu_j \,.
\end{equation}
Due to loss of coherence the phases are irrelevant.

The strongest change of flavors of the eigenstates~\eqref{eq:flavcont}
occurs when neutrinos cross the resonance layer centered at the
resonance density given by the resonance condition~\cite{ms1a, ms1b,
  ms2}:
\begin{equation}
  V_{e}(n^\text{res}) = \cos 2 \theta \frac{\Dmq}{2E} \,.
\end{equation}
The width of the layer is proportional to mixing: $n^\text{res} \tan
2\theta$.  The strongest change of flavor of whole the state is
realized when the initial density is much larger and final density is
much smaller than the resonance density. Resonance manifests itself
via dependence of $U^m_{ej}(n^0)$ on energy, and it corresponds to
maximal mixing.  The resonance condition is satisfied inside the Sun
for $E > 2~\text{MeV}$.


\subsection{From the Sun to the Earth}
\label{sec:evol-vacuum}

The wave functions (wave packets) of the eigenstates are determined by
processes of production of neutrinos.  Sizes of the wave packets are
different for different components of the solar neutrino spectrum
($pp$, \Nuc{7}{Be}, \Nuc{8}{B}, \textit{etc.}).  On the way from the
production point in the Sun to the Earth two effects happen: (i) the
wave packets (WP) of different eigenstates shift with respect to each
other and eventually separate in space due to different group
velocities; (ii) each WP spreads in space due to presence of different
momenta in it.

The first effect leads to loss of the propagation coherence (for low
energy neutrinos this happens already inside the Sun).
Restoration of coherence in a detector~\cite{Kiers:1995zj} would
require extremely long coherence time of detection process, or
equivalently, unachievable energy resolution: $\Delta E / E < l_\nu /
L_\text{Earth} \sim 2.5 \times 10^{-6} \, (E/10~\text{MeV})$, where
$L_\text{Earth}$ is the distance from the Sun to the Earth.

The spread is proportional to square of absolute value of mass, $m^2$.
It is much bigger than the original size of the packet for two heavier
neutrinos even for hierarchical spectrum.  Although the spread is
smaller than separation, and in any case, it does not affect the
coherence condition~\cite{kersten}.

Thus, incoherent fluxes of the mass eigenstates arrive at the Earth.
According to Eq.~\eqref{eq:nuesurf} their weights (admixtures) are
given by moduli squared of the mixing elements at the production point
$|U^m_{ej}(n_e^0) |^2$.  Therefore the probability to find $\nu_e$ in
the moment $t_E$ of the arrival equals
\begin{equation}
  \label{eq:patearth}
  P_{ee} = |\langle \nu_e | \nu (t_E) \rangle |^2
  = \sum_j |U^m_{ej}(n^0)|^2 |U_{ej}|^2 \,.
\end{equation}
In the standard parametrization of the mixing matrix
\begin{equation}
  \label{eq:standard}
  \begin{aligned}
    U^m_{e1} &= \cos \theta_{13}^m \cos \theta_{12}^m \,, \\
    U^m_{e2} &= \cos \theta_{13}^m \sin \theta_{12}^m \,, \\
    |U^m_{e3}| &= |\sin \theta_{13}^m| \,,
  \end{aligned}
\end{equation}
and in~\eqref{eq:patearth} the mixing angles in matter should be taken
in the production point: $\theta_{12}^m = \theta_{12}^m (n^0)$,
$\theta_{13}^m = \theta_{13}^m (n^0)$.  In terms of the mixing angles
the probability $P_{ee}$ equals
\begin{equation}
  \label{eq:nueday1}
  P_{ee} = c_{13}^2 c_{13}^{m2} P_2^{ad}
  + s_{13}^2 s_{13}^{m2} \,,
\end{equation}
where
\begin{align}
  \label{eq:adform}
  P_2^\text{ad} &= s_{12}^2 + \cos 2\theta_{12} \cos^2 \theta_{12}^m
  \\
  &= \frac{1}{2}(1 + \cos 2\theta_{12} \cos 2 \theta_{12}^m) \,.
\end{align}
The 1-2 mixing angle $\theta_{12}^m$ is determined by
\begin{equation}
  \cos 2 \theta_{12}^m =
  \frac{\cos 2\theta_{12} - c_{13}^2 \epsilon_{12}}{\sqrt{(\cos 2\theta_{12}
      - c_{13}^2 \epsilon_{12})^2 + \sin^2 2\theta_{12}}}
\end{equation}
with
\begin{equation}
  \label{eq:eps12}
  \epsilon_{12} \equiv \frac{2V_e E}{\Dmq_{21}} \,.
\end{equation}
The first term in~\eqref{eq:adform} gives the asymptotic ($E \to
\infty$) value of probability which corresponds to the non-oscillatory
transition, so that $P_{ee} \geq c_{13}^2 c_{13}^{m2} s_{12}^2$; the
second term describes effect of residual oscillations; the last term
in~\eqref{eq:nueday1} is the contribution of the decoupled third state
$\nu_3$.

The 1-3 mixing in matter in the production point can be estimated
as~\cite{Goswami:2004cn}
\begin{equation}
  \label{eq:13matter}
  \sin^2 \theta_{13}^m = \sin^2 \theta_{13} (1 + 2 \epsilon_{13})
  + \mathcal{O}(s_{13}^2 \epsilon_{13}^2, s_{13}^4 \epsilon_{13}) \,,
\end{equation}
where
\begin{equation}
  \label{eq:eps13}
  \epsilon_{13} \equiv \frac{2V_e (n_e^0) E}{\Dmq_{31}} \,.
\end{equation}
The correction in~\eqref{eq:13matter} can be as large as $12\%$.

Nature has selected the simplest (adiabatic) solution of the solar
neutrino problem. If the adiabaticity is broken, the probability would
acquire an additional term~\cite{Haxton:1986dm, Parke:1986jy}
\begin{equation}
  \Delta P_{ee} \approx
  -P_{12} c_{13}^2 c_{13}^{m 2} \cos 2\theta_{12}^m \cos 2\theta_{12} \,,
\end{equation}
where $P_{12}$ is the probability of transition between the
eigenstates during propagation~\cite{Haxton:1986dm, Parke:1986jy}.  If
initial density is above the resonance one, so that $\cos
2\theta_{12}^m < 0$, the correction is positive, which means that
adiabaticity violation weakens suppression of the original
$\nu_e$-flux.

Corrections to the leading order adiabatic approximation (adiabaticity
violation effect) equal
\begin{equation}
  \frac{\Delta P_{ee}}{P_{ee}}
  \approx \frac{\gamma^2 \cos 2\theta_{12}}{4 \sin^2 \theta_{12}} \,,
  \qquad
  \gamma = \frac{2 \dot \theta_m}{H_{2m} - H_{1m}} \,,
\end{equation}
where $\gamma$ is the adiabaticity parameter.  For $E = 10~\text{MeV}$
the correction is about $10^{-8}$~\cite{deHolanda:2004fd},
\textit{i.e.}, negligible.

For small mixing the jump probability $P_{12}$ is given by the
Landau-Zener formula~\cite{Parke:1986jy}, and the precise formula
valid also for large mixing angles has been obtained
in~\cite{Petcov:1987zj}.  Adiabaticity violation can be realized if,
\textit{e.g.}, hypothetical very light sterile neutrino exists, which
mixes very weakly with the electron neutrino (see
Sec.~\ref{sec:sterile}).


\subsection{Oscillations in matter of the Earth}
\label{sec:evol-earth}

Evolution in the Earth is more complicated than in the Sun~\cite{ms4,
  Bouchez:1986kb, Cribier:1986ak}.  Neutrino detectors are situated
underground and therefore oscillations in the Earth are present all
the times.  The oscillation lengths range from 10~km for low energy
$pp$-neutrinos to about 300~km for high energy \Nuc{8}{B}-neutrinos.
For high energies, oscillations in the Earth during the day can be
neglected, whereas for low energies the oscillations are present
during a part of day, but the effect is very small due to smallness of
mixing in matter.

Crossing the Earth surface the neutrino mass eigenstates split into
the eigenstates of Hamiltonian in matter of the Earth, $\nu_{k m}$,
\begin{equation}
  \nu_j \to \tilde{U}_{jk}^m \, \nu_{k m} \,,
\end{equation}
and start to oscillate.  Here $\tilde{U}^m$ is the mixing matrix of
the mass states in matter.  So, oscillations in the Earth are purely
matter effect.  Probability to detect the electron neutrino is given
by
\begin{equation}
  \label{eq:nuesurea}
  P_{ee} = \sum_j |U^m_{ej}(n^0_e)|^2 P_{je} \,,
\end{equation}
where $P_{je}$ are the probabilities of oscillation transitions $\nu_j
\to \nu_e$.  During the day $P_{je} \approx |U_{ej}|^2$.

The matter effect of the Earth on the 1-3 mixing is very small, so
that $P_{3e} \approx s_{13}^2$.
Therefore the unitarity condition, $\sum_j P_{je} = 1$, becomes
$P_{1e} + P_{2e} = 1 - s_{13}^2$.  With this and the
parametrization~\eqref{eq:standard} the Eq.~\eqref{eq:nuesurea} gives
\begin{equation}
  \label{eq:nuesurea2}
  P_{ee} = P_{1e}^E c_{13}^{m2} \cos 2\theta_{12}^m
  + c_{13}^2 c_{13}^{m2} \sin^2 \theta_{12}^m
  +  s_{13}^2 s_{13}^{m2}.
\end{equation}
So, the Earth matter effect is described by single oscillation
probability $P_{1e}^E$.
For solar neutrino energies the low density limit is realized when
\begin{equation}
  \epsilon_{12} =
  0.035 \left( \frac{E}{10~\text{MeV}} \right) \ll 1 \,.
\end{equation}
(Here $V_e$ is taken for the surface density).  In the lowest order in
$V_e (x)$ or $\epsilon_{12}$ and for arbitrary density profile the
probability equals $P_{1e}^E = c_{13}^2 c_{12}^2 - F_\text{reg}$,
where the regeneration factor is given by~\cite{Ioannisian:2004jk,
  Akhmedov:2004rq}
\begin{equation}
  \label{eq:prob-earth}
  F_\text{reg} =
  \frac{1}{2}c_{13}^4\sin^2 2\theta_{12}
  \int_0^L dx V_e (x) \sin \phi^m_{x \to L} \,.
\end{equation}
Here $\phi^m_{x \to L}$ is the phase acquired from a given point of
trajectory $x$ to a detector:
\begin{equation}
  \phi^m_{x \to L} = \int_x^L dy \Delta^m_{21} (y) \,,
\end{equation}
and the difference of eigenvalues equals
\begin{equation}
  \label{eq:diff-eigen}
  \Delta^m_{21} (y)
  = \frac{\Dmq_{21}}{2E}
  \sqrt{\left[ \cos 2\theta_{12} - c_{13}^2\epsilon_{12} (y) \right]^2
    + \sin^2 2\theta_{12}} \,.
\end{equation}

During the day, when effect of oscillations inside the Earth can be
neglected: $P_{1e} = U_{e1}^2 = c_{13}^2 c_{12}^2$.  Then according
to~\eqref{eq:nuesurea2} and~\eqref{eq:prob-earth} the difference of
probabilities with and without oscillations in the Earth (the
day-night asymmetry) equals
\begin{multline}
  \label{eq:diff-prob}
  P - P_0 =
  -c_{13}^2 \cos 2\theta_{12}^m F_\text{reg}
  \\
  = -\frac{1}{2}c_{13}^6 \sin^2 2\theta_{12}
  \cos 2\theta_{12}^m
  \int_0^L dx V_e (x) \sin \phi^m_{x \to L} \,.
\end{multline}
It is proportional to $c_{13}^6$ (see~\cite{Blennow:2003xw,
  Minakata:2010be}), so that non-zero 1-3 mixing reduces effect by
about $7 \%$.

Some insight into the results can be obtained in the constant density
approximation:
\begin{equation}
  \label{eq:prob-const1}
  F_\text{reg} =
  \sin^2 2\theta_{12} \left(\frac{c_{13}^4 V_e}{\Delta^m_{21}}\right)
  \sin^2 \frac{1}{2} \Delta^m_{21} L \,.
\end{equation}
Oscillations in the Earth reduce $P_{1e}^E$.  Consequently, for high
energy part of the spectrum with $\cos 2\theta_{12}^m < 0$ the effect
is positive, thus leading to regeneration of the $\nu_e$ flux, whereas
for low energies one has $\cos 2\theta_{12}^m > 0$, and the
oscillations in the Earth further suppress the $\nu_e$ flux.  The
regeneration effect approximately linearly increases with the neutrino
energy.

Equivalently, the result~\eqref{eq:diff-prob} can be obtained using
adiabatic perturbation theory~\cite{deHolanda:2004fd}.  Next order
approximation in $\epsilon_{21}$ has been obtained
in~\cite{Ioannisian:2004vv}.

Propagation in the Earth can be computed explicitly taking into
account that the matter density profile consists of several layers
with slowly changing density in which propagation is adiabatic and
density jumps at the borders of the layers where adiabaticity is
broken maximally.  The latter is accounted by matching conditions of
no flavor change.

Study of the Earth matter effect provides complete (integrated) check
of the solution of the solar neutrino problem, since all the phases of
evolution are involved.

The salient feature of this picture is that the third eigenstate
essentially decouples from evolution of rest of the system in all the
phases.  That is, any interference effect of $\nu_3$ or $\nu_{3m}$
with two other eigenstates is averaged out at the integration over
energy, or equivalently due to separation of the corresponding wave
packets.


\subsection{Averaging and attenuation}

Observable effects are determined by the $\nu_e$ survival probability
integrated over energy with resolution function of a detector, over
the kinematic distribution (in the case of $\nu -e$ scattering) and
over the energy profile of neutrino lines (\textit{e.g.}, the
\Nuc{7}{Be}-neutrino line).  This integration leads to the attenuation
effect~\cite{Ioannisian:2004jk} according to which a detector with the
energy resolution $\Delta E$ can not ``see'' remote structures of the
density profile for which distance to the detector is larger than the
attenuation length $\lambda_\text{att} \sim 1/\Delta E$.  In the core
due to larger density the oscillations proceed with larger
depth. However, this increase is not seen in boron neutrinos due to
the attenuation.  In contrast, for the \Nuc{7}{Be}-neutrinos the
energy resolution is given by the width of the line and
$\lambda_\text{att}$ turns out to be bigger than the distance to the
core.  So, detectors of \Nuc{7}{Be}-neutrinos can in principle ``see''
the core.

The probabilities should be averaged over the production region in the
Sun.  In the first approximation this can be accounted by the
effective initial densities $n_e^0 \to
\bar{n}_e^0$~\cite{deHolanda:2004fd}.


\subsection{Energy profile of the effect}

Flavor conversion is described by $P_{ee}(E, t)$~\eqref{eq:nuesurea2}
which depends on neutrino energy and time.  The time dependence is due
to oscillations in the Earth since the effect depends on the zenith
angle of trajectory of neutrino. The main dependence on energy is in
$\theta_{12}^m (n_0)$, and much weaker one is in $\theta_{13}^m (n_0)$
and $P_{1e}$.

\begin{figure}[t] 
  \includegraphics[width=\linewidth]{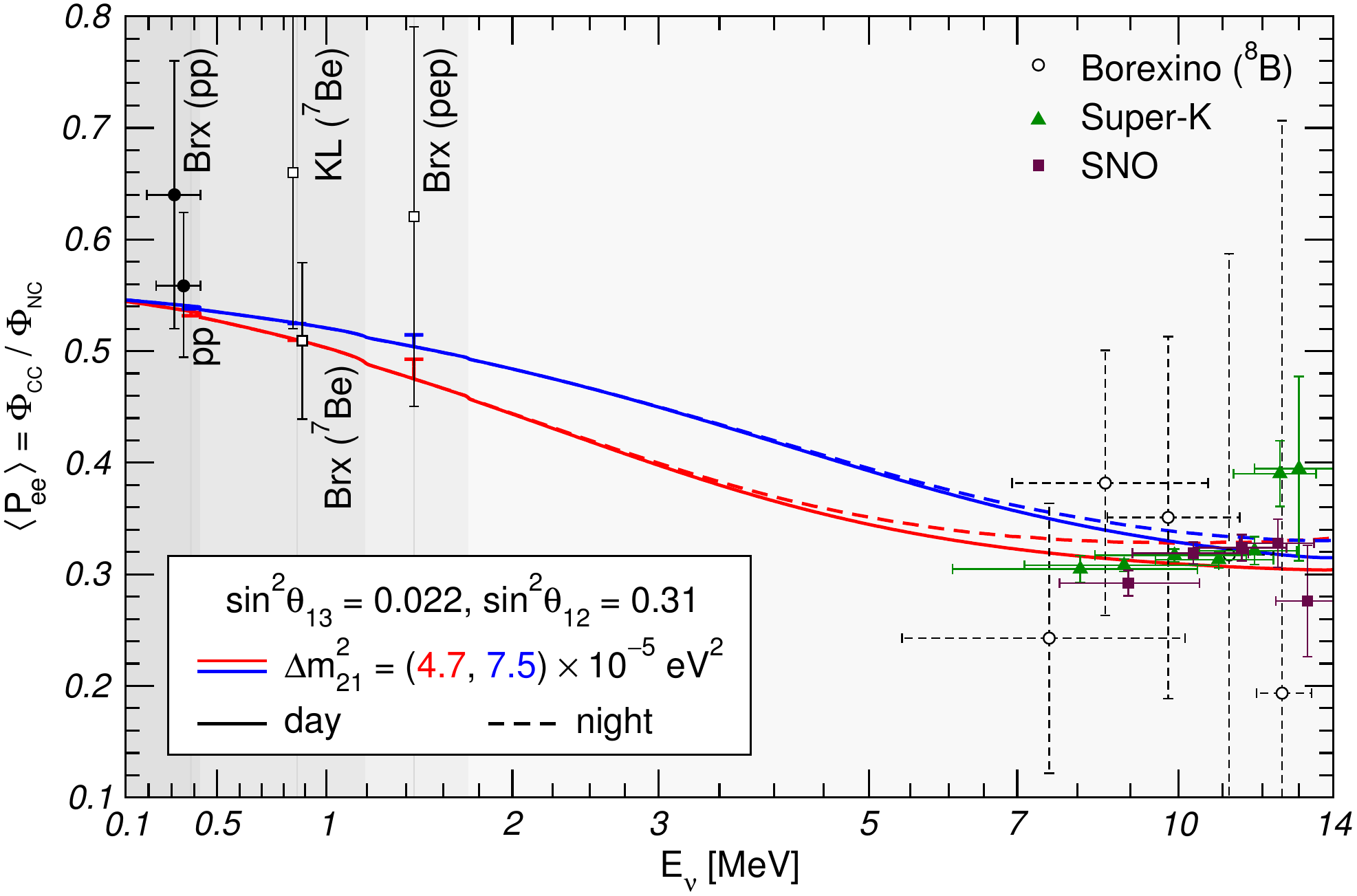}
  \caption{Dependence of the probabilities $P_{ee}$ integrated over
    the day and night time periods, for $\Dmq_{21}$ determined from
    the global fit of the solar neutrino data only (red) and from the
    global fit of all oscillation data (blue).  Also shown are the
    results from different experiments. We use abbreviations ``Brx''
    for Borexino and ``KL'' for KamLAND.}
  \label{fig:profile}
\end{figure}

Fig.~\ref{fig:profile} shows dependence of the probabilities
$P_{ee}(E)$ integrated over the day and the night times.  At low
energies neglecting the $\nu_e$ regeneration one has
\begin{equation}
  \label{eq:nuesur-vac}
  P_{ee} \approx
  c_{13}^4(1 - 0.5 \sin^2 2 \theta_{12}) - 0.5 c_{13}^6
  \cos 2\theta_{12} \sin^2 2\theta_{12} \epsilon_{12} \,.
\end{equation}
With decrease of energy: $P_{ee} \to P_{ee}^\text{vac}$.  For the best
fit value of the 1-2 mass splitting deviations of the
probability~\eqref{eq:nuesur-vac} from its vacuum values are $6\%$ for
the \Nuc{7}{Be}-neutrinos and $2\%$ for the $pp$-neutrinos with $E =
0.3~\text{MeV}$.

At high energies the matter effect dominates and
\begin{equation}
  \label{eq:peehigh}
  P_{ee} = c_{13}^4 \sin^2 \theta_{12} + c_{12}^2 F_\text{reg} + \frac{1}{4}
  \cos 2\theta_{12} \sin^2 2\theta_{12} \epsilon_{12}^{-2} \,.
\end{equation}

The intermediate energy region between the vacuum and matter dominated
limits is actually the region where the resonance turn on (turn
off). The middle of this region (before averaging) corresponds to the
MSW resonance at maximal densities in the Sun.  Value of $\theta_{12}$
determines sharpness of the transition, that is, the size of
transition region.  The larger $\theta_{12}$ the bigger the size of
the region.  Integration over the neutrino production region in the
Sun smears the transition, thus reducing the sensitivity to
$\theta_{12}$.

As follows from Fig.~\ref{fig:profile} almost all experimental points
are within $1\sigma$ from the prediction. Larger deviations can be
seen in the intermediate region.


\subsection{Scaling}

The conversion probability of solar neutrinos obeys certain scaling
which allows to understand various features of the LMA MSW solution as
well as effects of new physics. The survival probability averaged over
the oscillations on the way to the Earth (related to loss of
propagation coherence) is function three dimensionless parameters:
\begin{equation}
  \label{eq:proo}
  P_{ee} = P_{ee}(\epsilon_{12}, \epsilon_{13}, \phi_E) \,.
\end{equation}
Here
\begin{equation}
  \phi_E \approx \frac{\Dmq_{21} L}{2E}
\end{equation}
is the phase of oscillations in the Earth and $\epsilon_{12}$,
$\epsilon_{13}$ are defined in Eqs.~\eqref{eq:eps12}, \eqref{eq:eps13}
correspondingly.

Several important properties follow immediately:
\begin{enumerate}
\item The probability  is invariant with respect to rescaling
  \begin{equation}
    \label{eq:a-energy}
    \Dmq_{21} \to b \Dmq_{21} \,,
    \quad
    \Dmq_{31} \to b \Dmq_{31} \,,
    \quad
    E \to b E \,.
  \end{equation}

\item The adiabatic probability does not depend on distance and any
  spatial scale of the density profile. So, the only dependence on
  distance is in the phase $\phi_E$. If oscillations in the Earth are
  averaged, then whole the probability, $P_{ee} =
  P_{ee}(\epsilon_{12}, \epsilon_{13})$, is scale invariant. This
  happens for practically all values of the zenith angle.  In this
  case $P_{ee}$ is invariant with respect to rescaling
  \begin{equation}
    \label{eq:scalinga}
    \Dmq_{21} \to a \Dmq_{21} \,,
    \quad
    \Dmq_{31} \to a \Dmq_{31} \,,
    \quad
    V_e \to a V_e \,.
  \end{equation}
  In particular, if $a = -1$, $P_{ee}$ is invariant with respect to
  change of the signs of mass squared differences and potentials.
  Since the oscillation probability in the Earth (the regeneration
  factor) does not change under $\phi_E \to -\phi_E$, the invariance
  with respect to simultaneous change of signs of $\Dmq$ and potential
  (Eq.~\eqref{eq:scalinga} with $a = -1$) holds also for the
  non-averaged probability~\eqref{eq:proo}.

\item If $|\Dmq_{31}|$ is kept fixed, the scaling~\eqref{eq:scalinga}
  is broken by the 1-3 oscillations.

\item The dependence of the probability on $\epsilon_{13}$ is weak,
  and if neglected,
  \begin{equation}
    \label{eq:simpsc}
    P_{ee} \approx  P_{ee}(\epsilon_{12})
    = P_{ee} \left(\frac{2V_e E}{\Dmq_{21}} \right)
  \end{equation}
  depends on one combination of the parameters only.
\end{enumerate}
We will use these properties in the following discussion.


\section{Determination of the neutrino parameters}
\label{sec:param}

The conversion effect of the solar neutrinos depends mainly on $\sin^2
\theta_{12}$ and $\Dmq_{21}$. In the approximation $\sin^2 \theta_{13}
= 0$ the problem is reduced to $2\nu$ problem.
Due to low neutrino energies the 1-3 mixing, being small in vacuum, is
not enhanced substantially in matter.  Consequently, corrections to
the $2 \nu$ approximation are proportional to $\sin^2 \theta_{13}$.

With increase of experimental accuracy dependence of the probability
on the 1-3 mixing becomes visible. It is mainly via dependence of the
elements of PMNS matrix $U_{e1}$ and $U_{e2}$ on $\cos^2 \theta_{13}$.

Dependence of the probability on $\Dmq_{31}$ is via the matter
correction to the 1-3 mixing~\eqref{eq:13matter}. This correction is
about $0.3\%$ at 10~MeV, that is, an order of magnitude smaller than
correction due to the non-zero 1-3 mixing itself.

Similarly, the sensitivity of solar neutrinos to the 1-3 mass
hierarchy (the sign of $\Dmq_{31}$) is low.  According to
Eq.~\eqref{eq:13matter} in the case of inverted mass hierarchy the
correction to the $\sin^2\theta_{13}$ is negative.  Consequently, the
survival probability increases at high energies by about $0.5\%$ in
comparison with the NH case.

Solar neutrinos are insensitive to the 2-3 mixing.  The reason is that
only the electron neutrinos are produced in the Sun, and $\nu_\mu$ and
$\nu_\tau$ can not be distinguished at the detection.

Solar neutrino fluxes do not depend on the CP-violation phase
$\delta$~\cite{Minakata:1999ze}. Indeed, in the standard
parametrization $|U_{ei}|$ do not contain $\delta$. In matter the
propagation can be considered in the propagation basis,
$\nu_\text{prop}$, defined as $\nu_f = U_{23} \Gamma_\delta
\nu_\text{prop}$, where $U_{23}$ is the matrix of rotation in the
($\nu_\mu$, $\nu_\tau$) plane and $\Gamma_\delta \equiv \diag(1, 1,
e^{i\delta})$.  In this basis the CP phase is eliminated from
evolution, whereas $\nu_e$ is unchanged. As a result, the amplitude of
probability, $A_{ee}$, does not depend on $\delta$.


\subsection{The 1-2 mixing and mass splitting}

The angle $\theta_{12}$ determines the energy dependence of the
effect~\eqref{eq:nuesurea2} (shape of the energy profile) via
$\theta_{12}^m$ both in the Sun and the Earth.  The oscillation phase
is relevant only for oscillations in the Earth for a small range of
zenith angles near horizon. Also in the first approximation the
dependence of the probability on $\Dmq_{31}$ can be neglected. Then
the whole the picture is determined by $\Dmq_{21}$ in combination with
energy: $\Dmq_{21}/E$. This means that with change of $\Dmq_{21}$ the
profile shifts in the energy scale by the same amount without change
of its shape.  In particular, with decrease of $\Dmq_{21}$ the same
feature of the profile (\textit{e.g.}, the upturn) will show up at
lower energies.  Since dependence of the profile on $E$ is weak at
large and small energies, it is the position of the transition region
with respect to the solar neutrino spectrum that determines
$\Dmq_{21}$.

Also the $\nu_e$ regeneration effect in the Earth depends on
$\Dmq_{21}$: according to~\eqref{eq:prob-earth} $F_\text{reg} \propto
1/ \Dmq_{21}$.

\begin{figure}[t] 
  \includegraphics[width=\linewidth]{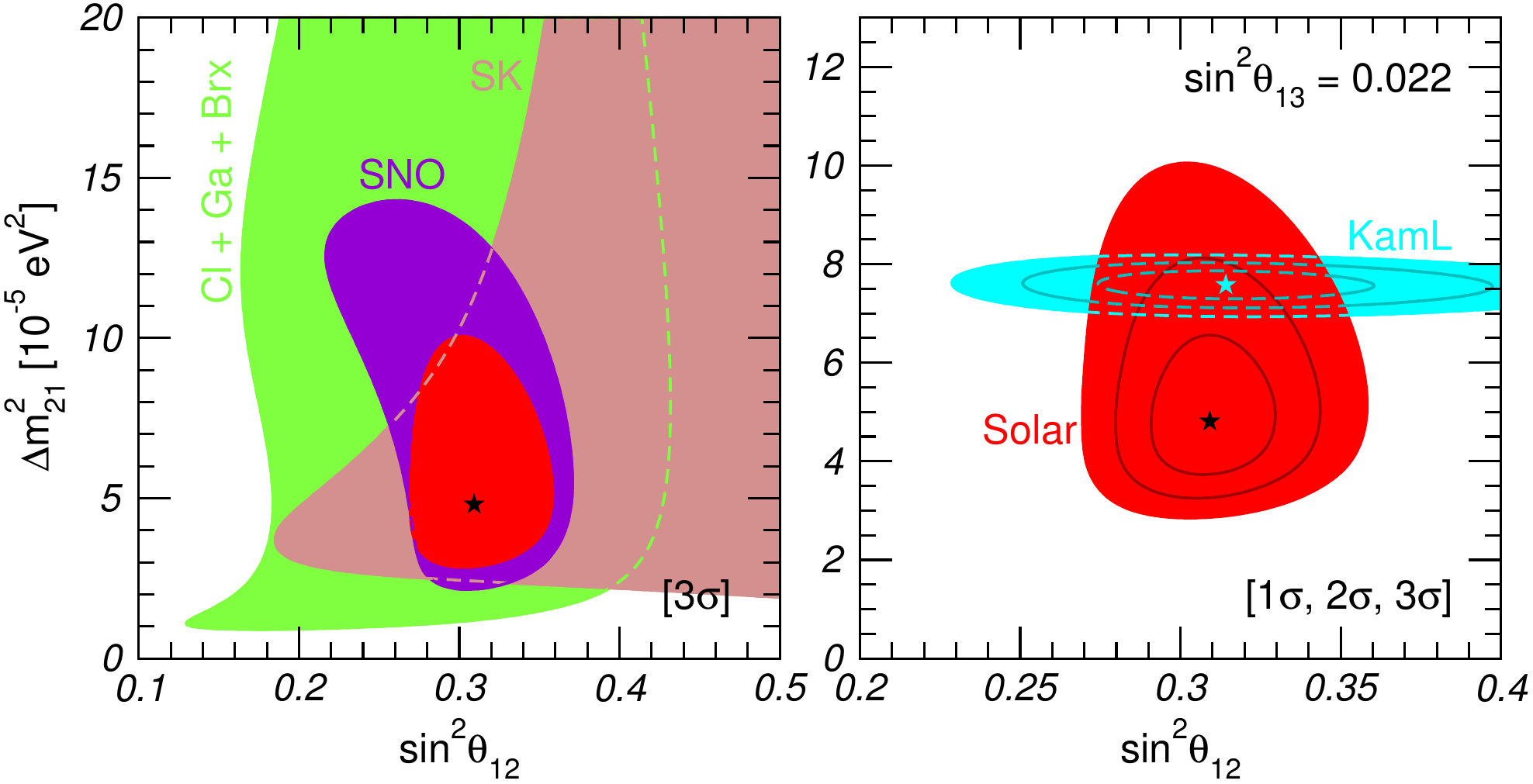}
  \caption{Allowed regions of $\theta_{12}$ and $\Dmq_{21}$ from the
    global fit of the solar neutrino data (red) as well as KamLAND
    (blue), for $\theta_{13}$ fixed to the best fit of the reactor
    experiments.  In the lefts panel shown are also regions restricted
    by individual experiments.}
  \label{fig:regions}
\end{figure}

In Fig.~\ref{fig:regions} we show result of the global fit of the
solar neutrino data in the ($\theta_{12}$, $\Dmq_{21}$) plane, for
$\theta_{13}$ fixed to the best fit value from the reactor
experiments.  In the left panel we show the regions restricted by
individual solar neutrino experiments, whereas in the right panel we
compare the solar and KamLAND allowed regions.  The preferred value of
$\theta_{12}$ from the analysis of solar data slightly increases as
$\theta_{13}$ increases.  Compared to the solar neutrino analysis
KamLAND gives about $2 \sigma$ larger $\Dmq_{21}$ but practically the
same value of $\theta_{12}$.


\subsection{The 1-2 mass ordering}
\label{sec:12order}

Solar neutrinos allow to fix the sign of $\Dmq_{21}$ for the standard
value of $V_e$.  The sign determines the resonance channel (neutrino
or antineutrino) and the mixing in matter.  The facts that due to
smallness of the 1-3 mixing the problem is reduced approximately to
the $2\nu$-problem and that suppression of signal averaged over the
oscillations at high energies is stronger than $1/2$, selects
$\Dmq_{21} > 0$.  That corresponds to the normal ordering (hierarchy)
when the electron flavor is mostly present in the lightest state.

For both signs of $\Dmq_{21}$ consideration and formulas are the same
and the only difference is the value of $\theta_{12}^m(n_e^0)$ in the
production point.
For high energies when density at production is much bigger than the
resonance one $n_e^0 \gg n^{res}(E)$, one has $\cos 2\theta_{12}^m
\approx -1$ ($+1$) for normal (inverted) ordering. Correspondingly,
the ratio of probabilities in the NH and IH cases equals $\tan^2
\theta_{12} \approx 1/2$.  Thus, for inverted ordering the suppression
would weaken with increase of energy.

According to Eq.~\eqref{eq:prob-const1} with change of the sign of
$\Dmq_{21}$ the Earth matter effect (regeneration factor) flip the
sign.


\subsection{The 1-3 mixing}

\begin{figure}[t] 
  \includegraphics[width=\linewidth]{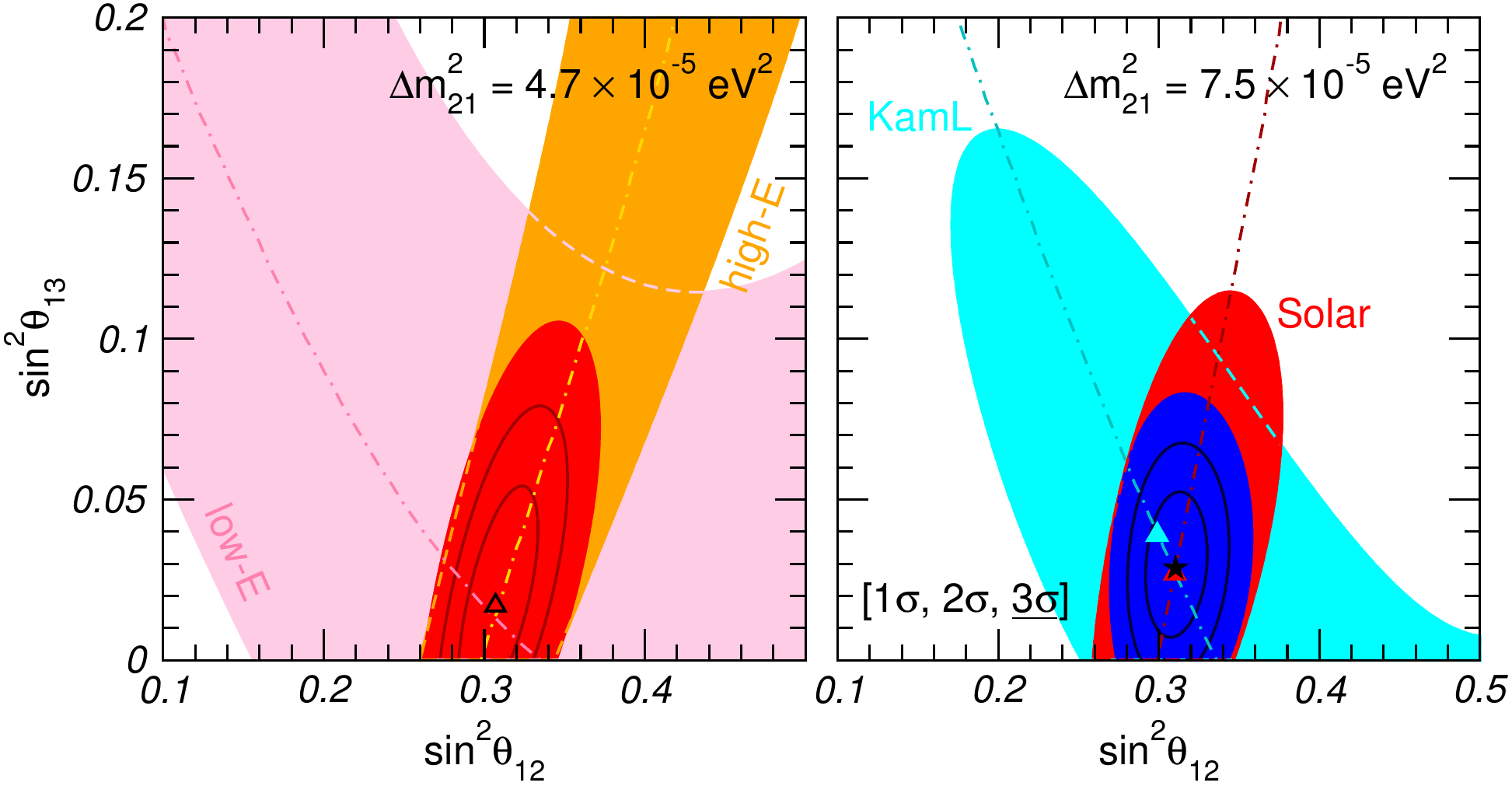}
  \caption{Allowed regions of mixing angles $\theta_{12}$ and
    $\theta_{13}$ from the global fit of the solar neutrino data only
    (left) and the solar plus KamLAND data (right).}
  \label{fig:tension}
\end{figure}

If $s_{13}^4$ terms in the probability~\eqref{eq:nueday1} are
neglected, dependence on the 1-3 appears as an overall normalization
which can be absorbed (at least partially) in uncertainties of
neutrino fluxes.  In contrast, degeneracy of the 1-2 and 1-3 mixings
is absent since $\sin^2 \theta_{12}$ and $\sin^2 \theta_{13}$ enter
the probability $P_{ee}$ in different combinations in the vacuum and
matter dominated energy regions.  According to~\eqref{eq:nuesur-vac}
and~\eqref{eq:peehigh} these combinations are
\begin{equation}
  \label{eq:s13lh}
  s_{13}^2 \approx \frac{1}{2} - \frac{P^h}{2 s_{12}^2} \,,
  \qquad
  s_{13}^2 \approx \frac{1}{2} - \frac{P^l}{2 - \sin^2 2\theta_{12}} \,.
\end{equation}
The SNO and SK results on the one hand side, and Borexino
(\Nuc{7}{Be}-, $pp$-neutrinos), and to a large extent Ga-Ge results on
the other depend on different combinations of angles $\theta_{12}$ and
$\theta_{13}$.  Fig.~\ref{fig:tension} shows the allowed region in the
plane $\sin^2 \theta_{12} - \sin^2 \theta_{13}$.
The left panel illustrates how the low and high energy data restrict
the allowed region. Also shown is the result of global fit of all
solar neutrino data which gives smaller $\Dmq_{21}$ and the best fit
value $\sin^2 \theta_{13} = 0.017$. The latter coincides with the
earlier result in Ref.~\cite{Goswami:2004cn}.

Instead of low energy solar neutrino data one can use the KamLAND
antineutrino result (vacuum oscillations with small matter
corrections).  Result of the combined fit of the solar and KamLAND
data in assumption of the CPT invariance is shown in
Fig.~\ref{fig:tension} (right).  KamLAND data shift the 1-3 mixing to
bigger value: $\sin^2 \theta_{13} = 0.028$.  The present solar
neutrino accuracy on $\theta_{13}$ is much worse than the one from the
reactor experiments, but it can be substantially improved in future by
SNO+, JUNO, Hyper-Kamiokande.


\subsection{Tests of theory of neutrino oscillations and conversion}
\label{sec:potential}


\subsubsection{Determination of the matter potential}

As discussed in the previous sections, the MSW effect plays central
role in the solution of the solar neutrino problem. It is therefore
important to experimentally verify all the aspects of this effect, and
in particular, value of matter potential.  To this end, we follow the
approach of Ref.~\cite{Fogli:2003vj} and allow for an overall
rescaling of the matter potential:
\begin{equation}
  V_e \to a_\textsc{msw} V_e \,.
\end{equation}
Note that such a modification of the matter term can be regarded as a
special case of non-standard neutrino
interactions~\cite{Gonzalez-Garcia:2013usa}, which we will describe in
detail in Sec.~\ref{sec:nsi}.

\begin{figure}[t] 
  \includegraphics[width=\linewidth]{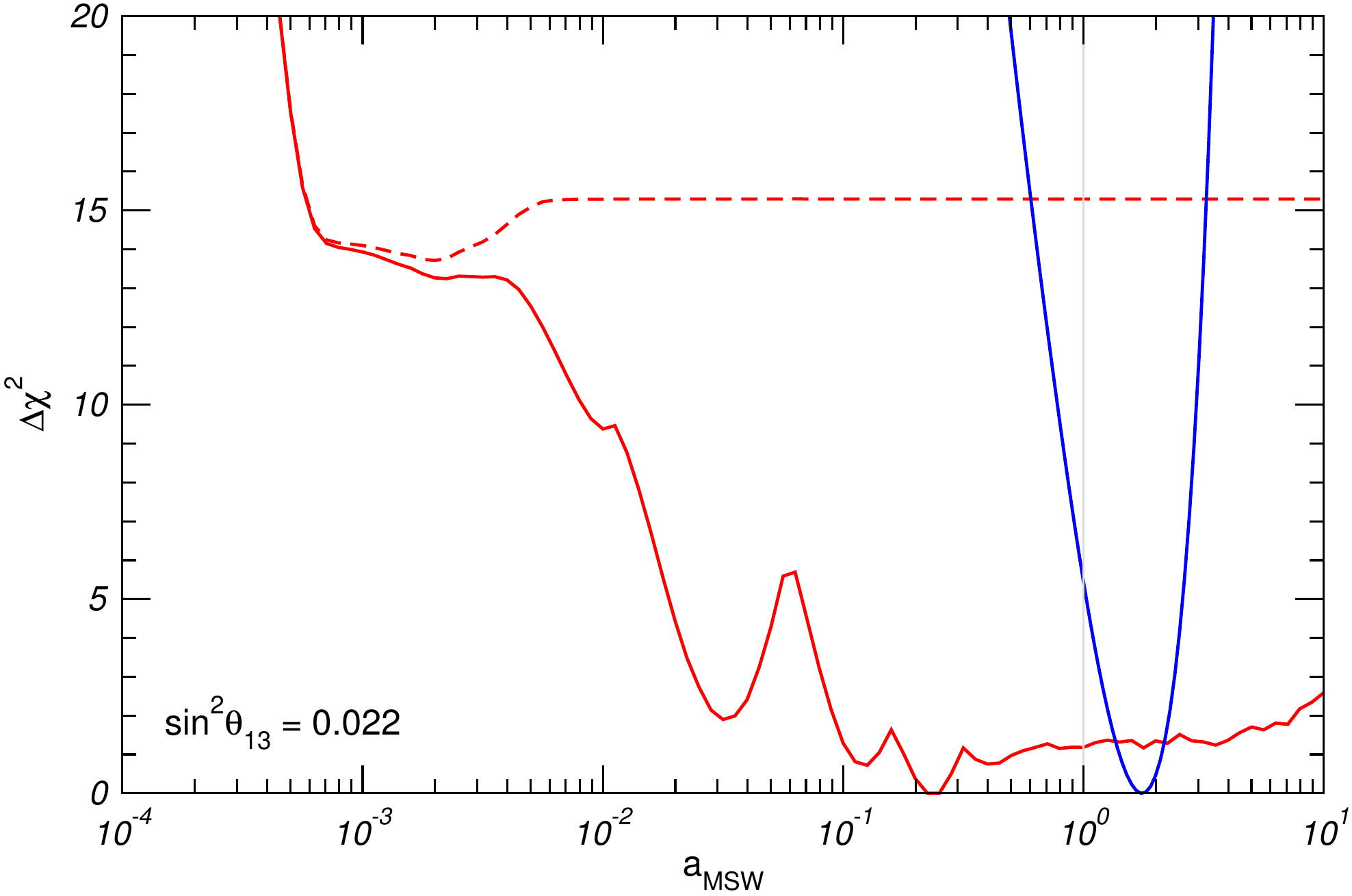}
  \caption{Determination of the matter potential from the solar (red)
    and solar+KamLAND (blue) neutrino data, as a function of the
    scaling parameter $a_\textsc{msw}$. We fix $\sin^2\theta_{13} =
    0.022$ and $\Dmq_{31} \to \infty$, and marginalize over
    $\Dmq_{21}$ and $\theta_{12}$. The dashed red line is obtained
    neglecting the Earth matter effect.}
  \label{fig:potential}
\end{figure}

In order to determine the preferred value and allowed range of
$a_\textsc{msw}$, we perform a fit of the solar neutrino data only
(Fig.~\ref{fig:potential}, red line) and a combined fit of the solar
and KamLAND neutrino data (blue line).  We fix for simplicity
$\sin^2\theta_{13} = 0.022$ but allow $\theta_{12}$ and $\Dmq_{21}$ to
vary freely.  We find $0.84 \le a_\textsc{msw} \le 3.08$ at the
$3\sigma$ level (see Fig.~\ref{fig:potential}, blue line), with
best-fit value $a_\textsc{msw} = 1.66$. The standard value
$a_\textsc{msw} = 1$ is well within the allowed region, although
slightly disfavored by the data ($\Delta\chi^2=5$). As we will see in
Sec.~\ref{sec:nsi}, this is related to the tension between solar and
KamLAND data in the determination of $\Dmq_{21}$, which can be
alleviated by a non-standard matter potential.

Inclusion of the KamLAND data is essential for determination of
$a_\textsc{msw}$. Indeed, as long as scaling~\eqref{eq:simpsc} is
realized ($P_{ee}$ depends only on the combination $V_e / \Dmq_{21}$),
a rescaling of the matter potential $V_e$ can be compensated by the
same rescaling of $\Dmq_{21}$.  This is clearly reflected by the
dashed red line in Fig.~\ref{fig:potential}, for which the Earth
matter effect has been ``switched off''.

According to Fig.~\ref{fig:potential}, scaling~\eqref{eq:simpsc} is
broken in the range $3 \times 10^{-3} \lesssim a_\textsc{msw} \lesssim
0.3$ by the effect of non-averaged oscillations in the Earth, which
for small $\Dmq_{21}$ depend on $L$.  This is seen as wiggles of the
red solid line.  Violation of scaling at $a_\textsc{msw} \lesssim 5
\times 10^{-3}$ corresponds to the breaking of adiabaticity in the
Sun, so that formula for $P_{ee}$ becomes invalid.  In this range
$l_\nu > 0.1 R_\text{sun}$.  The sharp increase of $\Delta\chi^2$ at
$a_\textsc{msw} \sim 5 \times 10^{-4}$ (very small $\Dmq_{21}$) occurs
when $l_\nu \sim R_\text{sun}$.

Thus, the solar neutrinos alone give only the lower bound
$a_\textsc{msw} > 5 \times 10^{-3}$, and can not fix $V_e$ and
$\Dmq_{21}$.  The precise matter-independent determination of
$\Dmq_{21}$ provided by KamLAND fixes the issue.

According to Fig.~\ref{fig:potential} at $a_\textsc{msw} \sim 1$ the
Earth matter effect reduces $\chi^2$ by $\Delta\chi^2 \sim 14$,
\textit{i.e.}, the effect is seen at about $4\sigma$, when all the
data (in particular, SNO) are included.


\subsubsection{Oscillation phase}

Results of global analysis of the solar neutrino data can be compared
with results from non-solar experiments which have different
environment, type of neutrino, energy, \textit{etc}. This comparison
provides important possibility to test the theory of oscillations and
to search for new physics.

As an example, suppose that for some geometrical reasons,
non-locality, \textit{etc.}, the phase in the oscillation formula
differs from the standard one by a factor $\xi$:
\begin{equation}
  \phi \to \xi \phi
\end{equation}
as it was advocated in some publications earlier.  This means that
$\Dmq_{21}$ extracted from the corresponding measurements would be
different by a factor $1/\xi$.  If the factor $\xi$ does not appear in
the Hamiltonian, one can establish existence of $\xi \neq 1$ using
$\Dmq_{21}$ from the adiabatic conversion result which does not depend
on the phase.  The fact that $\Dmq_{21}$ obtained from solar neutrino
data and KamLAND are close to each other allows to restrict $\xi$.

Comparing results from KamLAND measurement and solar neutrino
experiments one can search for effects of CPT violation, presence of
non-standard interactions, \textit{etc}.


\subsection{Status of the LMA MSW. Open issues}

LMA MSW gives good description of all existing solar neutrino data
with their present accuracy and no statistically significant deviation
are found.  The LMA solution reproduces all the observed features
including the directly measured $pp$-neutrino
flux~\cite{Bellini:2014uqa}, as well as the Day-night asymmetry: $(2 -
4)\%$ for the boron neutrinos~\cite{Renshaw:2013dzu} and the value
consistent with zero for \Nuc{7}{Be}-neutrinos~\cite{Bellini:2011yj}.
Pulls of different measurements with respect to predictions for the
best-fit values of oscillation parameters extracted from the solar
neutrino data can be seen in Fig.~\ref{fig:profile}.  There is certain
redundancy of measurements from different experiments which provides
consistency checks of the results.

Realization of the KamLAND experiment~\cite{Suzuki:2014woa} was
motivated by the solar neutrinos studies, namely by a possibility to
test the LMA solution.  Although historically by measuring $\Dmq_{21}$
KamLAND has uniquely selected the LMA solution, now the solar neutrino
experiments alone can do this due to new measurements by Borexino,
which validated the solution at low energies, and due to higher
accuracy of other results.

There are a few open issues which motivate further detailed studies.
Three following facts are most probably related:
\begin{enumerate}
\item The ``upturn'' of the spectrum (the ratio of the measured
  spectrum to the SSM one) towards low energies is not observed.
  According to the LMA solution the suppression should weaken with
  decrease of energy is not observed (see fig.~\ref{fig:profile}).
  The increase should be for all energies (if oscillations in the
  Earth are not included) but the strongest change is expected in the
  range $(2 - 6)~\text{MeV}$.  With oscillations in the Earth also the
  upturn towards high energies is expected.  No one experiment has
  showed the upturn.  The SNO experimental points even turn down at
  low energies.

\item For the best fit values from the global solar neutrino fit, one
  expects the D-N asymmetry $A_\textsc{dn} \equiv 2(N - D)/(N + D) =
  2.8\%$.  For values of $\Dmq_{21}$ from global fit of all
  oscillation data (dominated by KamLAND) the asymmetry equals
  $A_\textsc{dn} = 1.8\%$.  Super-Kamiokande gives larger value:
  $A_{DN} = (3.2 \pm 1.1 \pm 0.5) \%$~\cite{Renshaw:2013dzu}, and even
  larger asymmetry, $4.2 \%$, has been obtained from separate day and
  night measurements.  This can be simply statistical fluctuation.
  Especially in view of the observed energy and zenith angle
  dependencies of the asymmetry.

\item The 1-2 mass splitting extracted from the global fit of the
  solar neutrino data $\Dmq_{21} = (4.7^{+1.6}_{-1.1}) \times
  10^{-5}~\eVq$ is about $2 \sigma$ smaller than the value measured by
  KamLAND (antineutrino channel) as well as the global fit value of
  all oscillation data~\cite{Gonzalez-Garcia:2014bfa} $\Dmq_{21} =
  (7.50_{-0.17}^{+ 0.19}) \times 10^{-5}~\eVq$.  Notice that the bump
  in the spectrum of reactor antineutrinos at $(4 - 6)~\text{MeV}$
  uncovered recently~\cite{Seo:2014jza, Zhan:2015aha} leads to a
  decrease of $\Dmq_{21}$ extracted from KamLAND data by about $0.1
  \times 10^{-5}~\eVq$, and therefore to an insignificant reduction of
  the disagreement~\cite{Maltoni15}.\footnote{This shift in
    $\Dmq_{21}$ has been obtained by fitting the 2013 KamLAND data
    presented in~\cite{Gando:2013nba} with a reactor antineutrino
    spectrum modified according to the RENO near-detector measurement
    shown in Fig.~6 of Ref.~\cite{Seo:2014jza}, and is in good
    agreement with the result reported recently
    in~\cite{Capozzi:2016rtj}.}  With the decrease of $\Dmq_{21}$ the
  upturn and regeneration peak shift to lower energies, which leads to
  weaker distortion of the spectrum at low energies and larger D-N
  asymmetry.

\item The value of potential extracted from the solar neutrino data is
  larger by a factor 1.6 than the standard potential. This is directly
  related to difference of $\Dmq_{21}$ extracted from the solar and
  KamLAND data (Sec.~\ref{sec:potential}).
\end{enumerate}

Solar neutrino studies motivated calibration experiments with
radiative sources. The latter led to Gallium anomaly --~about
$2\sigma$ deficit of signal which implies new physics unrelated to the
solar neutrinos (and has value by itself). Impact of the Gallium
calibration on results of solar neutrino experiments is not strong if
it is related to cross-section uncertainties in the energy range of
calibration sources.


\subsection{Theoretical and phenomenological implications}

Measured oscillation parameters have important implications for
fundamental theory, though there is no unique interpretation.  The
observed 1-2 mixing is large but not maximal.  The deviation of the
measured value $\sin^2 \theta_{12} = 0.304^{+0.013}_{-0.012}$ is about
$15 \sigma$ below $\sin^2 \theta_{12} = 0.5$ and it is substantially
larger than $\sin^2 \theta_C = 0.050$. The value of $\sin^2
\theta_{12}$ is in between of $\sin^2(\pi/4 - \theta_C) = 0.281$ and
$\sin^2 \theta_{12} = 1/3$, where the first number corresponds to the
Quark Lepton Complementarity (QLC)~\cite{Minakata:2004xt} and the
second one to the Tri-bimaximal (TBM) mixing~\cite{Harrison:2002er}.
In turn, QLC implies a kind of quark-lepton unification (symmetry),
and probably, the Grand Unification (GU).  TBM indicates toward
geometric origins of mixing and certain flavor symmetry which is
realized in the residual symmetries approach~\cite{Lam:2008sh,
  Lam:2008rs, Grimus:2009pg}.

Measured value of $\Dmq_{21}$ gives the lower bound on the mass $m_2
\geq \sqrt{\Dmq_{21}} = 0.007~\text{eV}$.  Comparing $\Dmq_{21}$ with
$\Dmq_{31}$ one finds that neutrinos have the weakest mass hierarchy
(if any) among all other leptons and quarks in the case of normal mass
ordering: $m_2/m_3 \geq \sqrt{\Dmq_{21}/\Dmq_{31}}$ $= 0.18$.  In the
case of inverted mass hierarchy $\Dmq_{21}$ determines degeneracy of
two heavy mass states $\Delta m /m_2 \geq \Dmq_{21} / 2\Dmq_{31} = 1.5
\times 10^{-2}$, which implies certain flavor symmetry. If neutrinos
are Majorana particles, their masses fix the effective scale of new
physics responsible for the neutrino mass generation. For the $D = 5$
Weinberg operator generating such a mass, we obtain the value new
physics scale $\Lambda = v_\text{EW}^2/\sqrt{\Dmq_{21}} \sim
10^{16}~\text{GeV}$ which coincides essentially with the GU scale.

There is a number of phenomenological consequences of the solar
neutrino results:
\begin{enumerate}
\item Supernova (SN) neutrinos: in outer regions of a collapsing star
  the MSW conversion produces significant flavor changes of
  fluxes. The conversion occurs in the adiabatic regime.  Due to the
  1-2 mass splitting and mixing the SN neutrinos oscillate in the
  matter of the Earth leading to the observable effects (see,
  \textit{e.g.}, Ref.~\cite{Dighe:1999bi}).

\item The Early Universe: equilibration of the lepton asymmetries in
  different flavors occurs due to oscillations with large
  mixings~\cite{Lunardini:2000fy, Dolgov:2002ab}.

\item Neutrinoless double beta decay: contribution from the second
  mass state to the effective Majorana mass of the electron neutrino
  gives the dominant contribution in the case of normal mass
  hierarchy: $m_{ee}^{(2)} \approx \sin^2 \theta_{12}
  \sqrt{\Dmq_{21}}$ $ = (2 - 3)~\text{meV}$. In the case of inverted
  hierarchy $m_{ee} \approx m_{ee}^{(1,2)} \propto c_{13}^2 |\cos^2
  \theta_{12} + \sin^2 \theta_{12} e^{i \phi}|$, and numerically
  $m_{ee}^{(1,2)} \approx (18 - 50)$~meV depending on value of the
  Majorana phase $\phi$.

\item CP-violation effects are proportional to $\sin 2\theta_{12}$,
  and $\Dmq_{21}$ determines the $L/E$ scale for oscillation
  experiments which are sensitive to the CP phase.
\end{enumerate}


\section{Solar neutrinos and physics beyond $3\nu$ framework}
\label{sec:bsm}

Apart from masses and mixing a number of non-standard neutrino
properties have been considered which lead to new effects in
propagation of neutrinos, and consequently, to suppression of the
solar $\nu_e$ flux. After establishing LMA MSW solution, these effects
can show up as subleading effects. Their searches in solar neutrinos
allow to put bounds on standard neutrino properties.

As far as propagation is concerned, effects of new physics can be
described by adding new terms, $H_\textsc{np}$, to the Hamiltonian in
Eq.~\eqref{eq:evolution}: $H \to H_\textsc{lma} + H_\textsc{np}$. In
the following sections we will use the approximation of the third mass
dominance, $\Dmq_{31} \to \infty$ or decoupling of $\nu_3$, according
to which the evolution of 3 neutrinos (in certain basis) is reduced to
the evolution of a $2\nu$ system described by an effective Hamiltonian
$H^{(2)} = H_\textsc{lma}^{(2)} + H_\textsc{np}^{(2)}$ with
\begin{equation}
  \label{eq:efham}
  H_\textsc{lma}^{(2)}
  = \frac{\Dmq_{21}}{4 E}
  \begin{pmatrix}
    -\cos 2\theta_{12} & \sin 2\theta_{12} \\
    \hphantom{+} \sin 2\theta_{12} & \cos 2\theta_{12}
  \end{pmatrix}
  + \sqrt{2} G_F n_e
  \begin{pmatrix}
    c_{13}^2 & 0 \\
    0 & 0
  \end{pmatrix}.
\end{equation}

In addition, new physics can affect production and interactions of
neutrinos, which we will discuss separately for each specific case.


\subsection{Sterile neutrinos}
\label{sec:sterile}

Sterile neutrinos, singlets of the standard model symmetry group, can
manifest themselves through mixing with ordinary neutrinos, with
non-trivial implications for the oscillation patterns.
The most general neutrino mass matrix which generates such a mixing
with $n$ sterile neutrinos has in the basis $\nu_f = (\nu_e, \nu_\mu,
\nu_\tau, \nu_{s1}, \dots, \nu_{s n})^T$ a form
\begin{equation}
  \label{eq:seesaw}
  \mathcal{M}_\nu =
  \begin{pmatrix}
    0 & m_D \\
    m_D^T & m_N
  \end{pmatrix},
\end{equation}
where $m_D$ is a generic $3\times n$ matrix and $m_N$ is a $n \times
n$ symmetric matrix.  Then the Hamiltonian in vacuum equals $H_0 =
\mathcal{M}_\nu \mathcal{M}_\nu^{\dagger}/2E$.  The diagonal matrix of
the matter potentials appearing in~\eqref{eq:evolution} is now $V =
\diag(V_e + V_n, V_n, V_n, 0, \dots, 0)$ with $V_n = -(1/\sqrt{2}) G_F
n_n$ and $n_n$ being the number density of neutrons.

From a phenomenological point of view, three different regimes can be
identified, depending on whether the mass-squared splitting involving
the sterile neutrinos, $\Dmq_\text{as}$, is much smaller, comparable,
or much larger than $\Dmq_{21}$.


\subsubsection{$\Dmq_\text{as} \ll \Dmq_{21}$: quasi-Dirac case}

This was realized in the ``Just-so'' solution of the solar neutrino
problem.  For very small Majorana masses, $|m_N| \ll |m_D|$, the
eigenvalues of Eq.~\eqref{eq:seesaw} form pairs of almost degenerate
states. This situation is referred as ``quasi-Dirac'' limit.  The
presence of extra mass-squared splittings can distort the neutrino
oscillation patterns, and due to very big baseline (the Sun-Earth
distance) the solar neutrino experiments have high sensitivity to
\emph{small} values of $|m_N|$.
For $\Dmq_\text{as} < 10^{-9}~\eVq$ the evolution inside the Sun is
practically unaffected by the presence of the sterile neutrinos (the
oscillations into sterile neutrinos are suppressed in matter), and the
main effect is due to vacuum oscillations $\nu_a \to \nu_s$ on the way
from the Sun to the Earth.

In order not to spoil the accurate description of solar oscillations
data the Majorana mass should satisfy the upper bound $|m_N| \lesssim
10^{-9}~\text{eV}$ (normal ordering) or $|m_N| \lesssim
10^{-10}~\text{eV}$ (inverted ordering)~\cite{deGouvea:2009fp,
  Donini:2011jh}.  The bound corresponds to the maximal active sterile
mixing and $\Dmq_\text{as} \approx 2 m_D m_N = (10^{-11} -
10^{-9})~\eVq$.


\subsubsection{$\Dmq_\text{as} \lesssim \Dmq_{21}$}

Originally, this possibility has been motivated by relatively low
$Ar$- production rate in the Homestake experiment and absense of
spectral upturn~\cite{deHolanda:2003tx, deHolanda:2010am}.  An extra
sterile state has been added to the three active ones, $\nu_f =
(\nu_s, \nu_e, \nu_\mu, \nu_\tau)^T$, and correspondingly, new mass
state $\nu_0$: $\nu_{m} = (\nu_0, \nu_1, \nu_2, \nu_3)^T$. The $4
\times 4$ mixing matrix is parametrized as $U = U_\text{PMNS}\,
U_{01}(\alpha)$~\cite{deHolanda:2003tx}.  The value of new mixing
angle is assumed to be very small: $\sin^2 2\alpha \sim 10^{-3}$ and
the new mass splitting equals $ \Dmq_{01} \sim 0.2 \Dmq_{21}~\eVq$.
The diagonal matrix of the matter potentials in the flavor basis is $V
= \diag(0, V_e + V_n, V_n, V_n)$.

\begin{figure}[t] 
  \includegraphics[width=\linewidth]{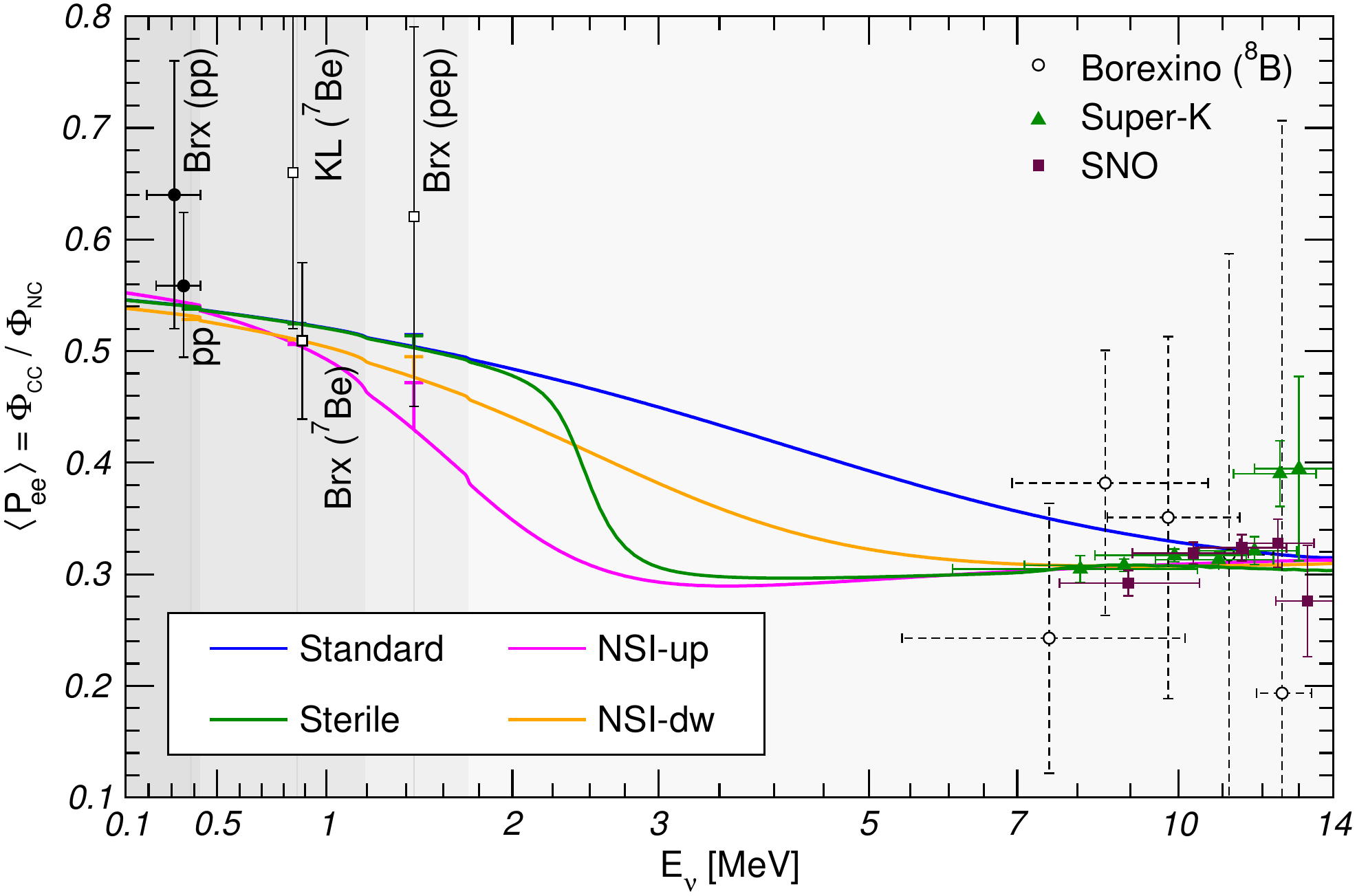}
  \caption{Impact of new physics on solar neutrino survival
    probability. We fix $\sin^2\theta_{13} = 0.022$. We plot standard
    oscillations ($\sin^2\theta_{12} = 0.31$, $\Dmq_{21} = 7.5\times
    10^{-5}~\eVq$), one extra sterile neutrino ($\sin^2\theta_{12} =
    0.31$, $\Dmq_{21} = 7.5\times 10^{-5}~\eVq$, $\sin^22\alpha =
    0.005$, $\Dmq_{01} = 1.2\times 10^{-5}~\eVq$) and non-standard
    interactions with up-type ($\sin^2\theta_{12} = 0.30$, $\Dmq_{21}
    = 7.25\times 10^{-5}~\eVq$, $\Eps_D^u = -0.22$, $\Eps_N^u =
    -0.30$) and down-type ($\sin^2\theta_{12} = 0.32$, $\Dmq_{21} =
    7.35\times 10^{-5}~\eVq$, $\Eps_D^d = -0.12$, $\Eps_N^d = -0.16$)
    quarks. See text for details. We use abbreviations ``Brx'' for
    Borexino and ``KL'' for KamLAND.}
  \label{fig:newphys}
\end{figure}

In such a model, the neutrinos propagating inside the Sun encounter
\emph{two} resonances: one is associated with the 1-2 mass splitting,
as in the standard case, and another one with the 0-1 mass splitting.
With parameters $\alpha$ and $\Dmq_{01}$ defined above the new
resonance modifies the survival probability leading to the dip at the
intermediate energies, $E \sim 3~\text{MeV}$, thus suppressing the
upturn (see Fig.~\ref{fig:newphys}). This alleviates the tension
between solar and KamLAND data.


\subsubsection{$\Dmq_\text{as} \gg \Dmq_{21}$}

In this limit (see Refs.~\cite{Palazzo:2011rj, Kopp:2013vaa} for
latest discussions) all the $\Dmq_{ij}$ other than $\Dmq_{21}$ can be
assumed to be infinite, and in certain propagation basis the neutrino
evolution is described by the sum of $H_\textsc{lma}^{(2)}$ in
Eq.~\eqref{eq:efham} and
\begin{equation}
  \label{eq:stereff}
  H_\textsc{np}^{(2)} = \sqrt{2} G_F \frac{n_n}{4}
  \begin{pmatrix}
    -\xi_D & ~\xi_N e^{-i\delta_{12}}
    \\
    \hphantom{-}\xi_N e^{i\delta_{12}} & ~\xi_D
  \end{pmatrix},
\end{equation}
where $\xi_D$, $\xi_N$ are combinations of the mixing matrix elements
$U_{\alpha i}$ (explicit expressions can be found in App.~C of
Ref.~\cite{Kopp:2013vaa}).

The new physics term~\eqref{eq:stereff} proportional to $V_n$, is
induced by the decoupling of heavy neutrino states.  In general, the
matter term~\eqref{eq:stereff} and the usual one with $V_e$ do not
commute with each other as well as with the vacuum term.  The phase
$\delta_{12}$ appearing in $H_\textsc{np}^{(2)}$ originates from the
phases of the general $(3+n)$ mixing matrix, and it cannot be
eliminated by a redefinition of the fields.  This phase does not
produce CP-violation asymmetry but affects neutrino propagation in
matter.

The relevant conversion probabilities can be written as
\begin{equation}
  \label{eq:sterprob}
  \begin{aligned}
    P_{ee} &= \tilde{C}_e - \eta_e^2 P_\text{osc}^{(2)} \,,
    \\
    P_{ae} &= \tilde{C}_a - \eta_e \left(
    \xi_D P_\text{osc}^{(2)} + \xi_N P_\text{int}^{(2)} \right),
  \end{aligned}
\end{equation}
where $P_\text{osc}^{(2)} \equiv |S_{21}^{(2)}|^2$ and
$P_\text{int}^{(2)} \equiv \Re\big( S_{11}^{(2)} S_{21}^{(2)\star}
\big)$ and the matrix $S^{(2)}$ is the solution of the evolution
equation with the effective Hamiltonian $H^{(2)}$.  The coefficients
$\tilde{C}_e$, $\tilde{C}_a$, and $\eta_e$ are functions of $U_{\alpha
  i}$~\cite{Kopp:2013vaa}.  The formulas~\eqref{eq:sterprob} are valid
for any number of sterile neutrinos.
Sterile neutrinos affect the oscillation probabilities in two
different ways:
\begin{enumerate}[label=(\arabic*)]
\item the mixing of $\nu_e$ with the ``heavy'' states leads to a
  suppression of the energy-dependent part of the conversion
  probabilities, in analogy with $\theta_{13}$ effects in the standard
  case;

\item the mixing of the sterile states with $\nu_{1,2}$ leads to
  overall disappearance of active neutrinos, so that
  $P_{ee} + P_{\mu e} + P_{\tau e} \ne 1$.
\end{enumerate}
Phenomenologically, the most relevant effect is the second one, since
the precise NC measurement performed by SNO confirms that the total
flux of active neutrinos from the Sun is compatible with the
expectations of the Standard Solar Model. Hence the fraction of
sterile neutrinos which can be produced in solar neutrino oscillations
is limited by the precision of the solar flux predictions, in
particular of the Boron flux. An updated fit of the solar and KamLAND
data in the context of (3+1) oscillations, with the simplifying
assumption $U_{e3} = U_{e4} = 0$, yields $|U_{s1}|^2 + |U_{s2}|^2 <
0.1$ at the 95\% CL.

Concerning the first effect, the mixing of $\nu_e$ with ``heavy''
eigenstates has similar implications as in the standard case except
that now there are ``more'' heavy states.  This allows to put a bound
on $\eta_e$ which is very similar to the one on $|U_{e3}|^2$ in $3\nu$
one, but instead of being interpreted as a bound on $|U_{e3}|^2$ it
becomes a bound on the sum $\sum_{i\ge 3} |U_{ei}|^2$. For example,
for (3+1) models a bound $|U_{e3}|^2 + |U_{e4}|^2 < 0.077$ at 95\% CL
can be derived from the analysis of solar and KamLAND data, as shown
in Ref.~\cite{Kopp:2013vaa}.  Additional bounds have been obtained by
Borexino~\cite{Bellini:2013uui}.


\subsection{Non-standard interactions}
\label{sec:nsi}

In the presence of physics Beyond the Standard Model, new interactions
may arise between neutrinos and matter. They can lead to effective
four-fermion operators of the form
\begin{equation}
  \mathcal{L}_\text{NSI} =
  - 2\sqrt{2} G_F \Eps_{\alpha\beta}^{fP}
  (\bar\nu_{\alpha} \gamma^\mu \nu_{\beta})
  (\bar{f} \gamma_\mu P f) \,,
\end{equation}
where $f$ denotes a charged fermion, $P \in \{L,R\}$ are the left and
right projection operators and $\Eps_{\alpha\beta}^{fP}$ parametrize
the strength of the non-standard interactions.

The non-standard interactions (NSI) were introduced to obtain
oscillations~\cite{w1a, w1b} or MSW conversion~\cite{Roulet:1991sm,
  Guzzo:1991hi} without neutrino masses.  NSI could provide an
alternative solution of the solar neutrino problem.  They can modify
the LMA MSW solution, and inversely, be restricted by solar
neutrinos~\cite{Friedland:2004pp, Palazzo:2009rb, Bolanos:2008km,
  Miranda:2004nb, Guzzo:2004ue}.

NSI affect neutrino propagation: the matter term, $V$, in the
evolution equation~\eqref{eq:evolution} includes an extra contribution
from NSI
\begin{equation}
  V_{\alpha\beta} = V_e \delta_{\alpha e} \delta_{\beta e}
  + \sqrt{2} G_F \sum_{f} n_f \, \Eps_{\alpha\beta}^f ,
\end{equation}
where $\Eps_{\alpha\beta}^f = \Eps_{\alpha\beta}^{fL} +
\Eps_{\alpha\beta}^{fR}$. Hermiticity requires that
$\Eps_{\beta\alpha}^f = \Eps_{\alpha\beta}^{f*}$, so that the diagonal
entries $\Eps_{\alpha\alpha}^f$ must be real.
The new physics part of the Hamiltonian equals
\begin{equation}
  \label{eq:nsieff}
  H_\textsc{np}^{(2)}
  = \sqrt{2} G_F  \sum_f n_f
  \begin{pmatrix}
    -\Eps_D^{f\hphantom{*}} & \Eps_N^f \\
    \hphantom{+} \Eps_N^{f*} & \Eps_D^f
  \end{pmatrix},
\end{equation}
where $\Eps_D^f$ and $\Eps_N^f$ are linear combinations of the
original parameters, $\Eps_{\alpha\beta}^f$, and their explicit
expressions can be found in Ref.~\cite{Gonzalez-Garcia:2013usa}.

Neglecting matter effect on the 1-3 mixing one obtains that the
$\nu_e$ probability $P_{ee}$~ can be written as $P_{ee} = c_{13}^4
P_\text{surv}^{(2)} + s_{13}^4$, where $P_\text{surv}^{(2)} \equiv
|S_{11}^{(2)}|^2$ should be calculated using the Hamiltonian
$H^{(2)}$.

In the specific case of NSI with \emph{electrons} ($f=e$) both the
standard and the non-standard (Eq.~\eqref{eq:nsieff}) terms scale with
the same matter density profile $n_e$.  This implies that large enough
positive value of $\Eps_D^e$ can ``flip the sign'' of the matter term,
so that the resonance will be realized for the inverted 1-2 hierarchy,
$\Dmq_{21} < 0$, in contrast to the usual case.  There is therefore an
unresolvable degeneracy between the sign of $\Dmq_{21}$ and that of
the matter potential: only their \emph{relative} sign can be
determined by oscillation experiments.  For NSI with up-quarks ($f=u$)
or down-quarks ($f=d$) this ambiguity is only approximate, however
present data are unable to resolve it. As a consequence, in the
presence of NSI the sign of $\Dmq_{21}$ can no longer be determined
uniquely.

\begin{figure}[t] 
  \includegraphics[width=\linewidth]{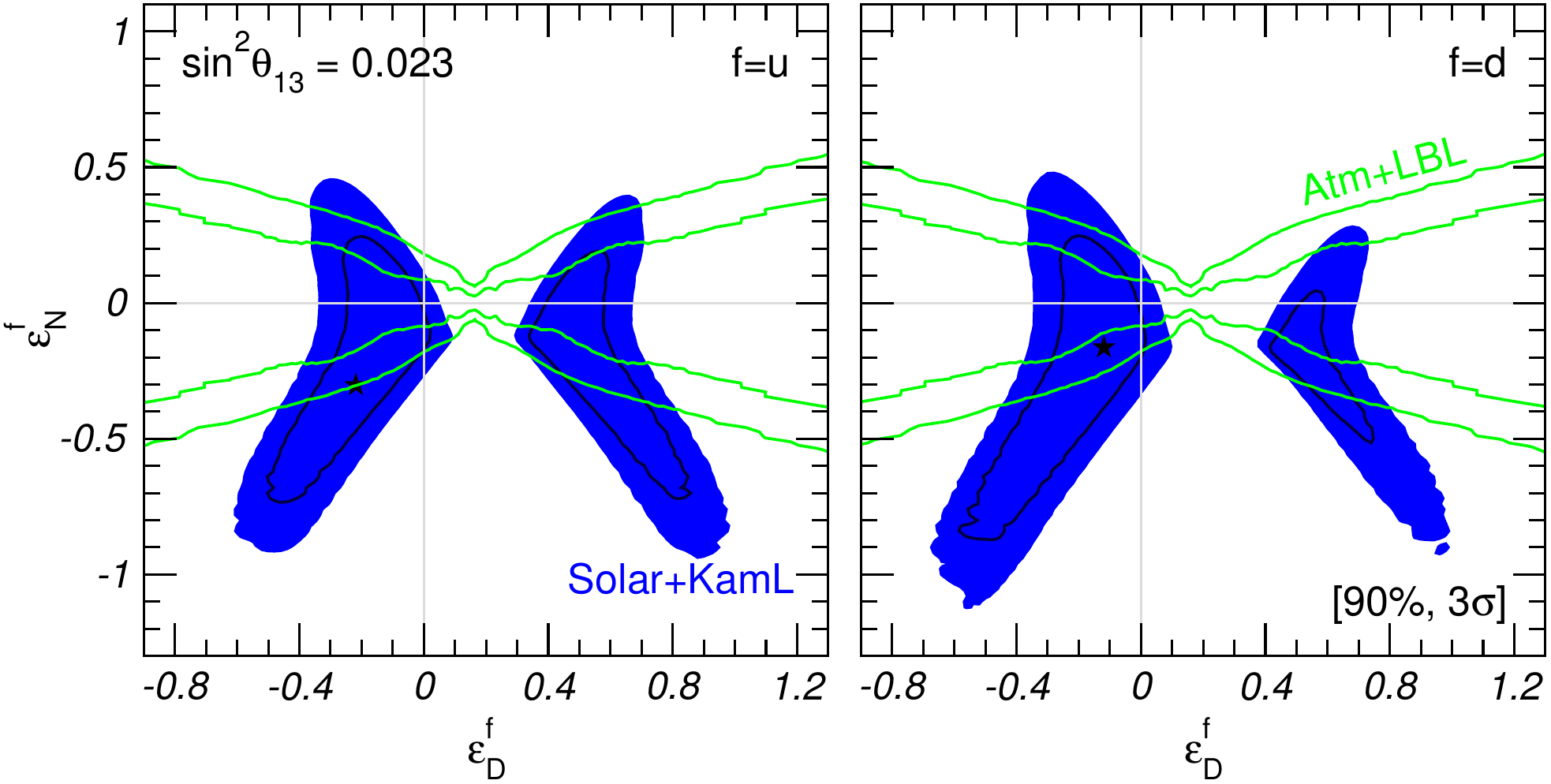}
  \caption{Allowed regions of parameters of the non-standard neutrino
    interaction (see text).}
  \label{fig:nonstand}
\end{figure}

With new interactions the evolution inside the Sun is still adiabatic,
and so the results are determined by the mixing at the production
point.  The latter is affected by NSI, and now also the off-diagonal
elements of the Hamiltonian depend on matter potential. This means
that at large values of the potential mixing is not suppressed: in
asymptotics, $E \to \infty$, one has
\begin{equation}
  \label{eq:costm}
  \cos 2 \theta_m  \approx
  -\frac{c_{13}^2 - 2r_f \Eps_D^f}{\sqrt{(2r_f \Eps_D^f - c_{13}^2)^2
      + (2 r_f \Eps_N^f)^2}},
\end{equation}
where $r_f \equiv n_f/n_e$.  According to Eq.~\eqref{eq:costm}, $\cos
2 \theta_m > -1$, and therefore suppression at high energies for the
same vacuum mixing is always weaker than without NSI. The mixing at
the exit from the Sun coincides with the vacuum mixing.  In the limit
of very low energies $P_{ee}$ approaches the vacuum value as in the
standard case.  The strongest modification appears in the intermediate
energy region.  In general, the probability is given
by~\eqref{eq:nueday1} with
\begin{equation}
  \label{eq:asymptot}
  \cos 2 \theta_m \approx
  -\frac{A}{\sqrt{A^2 + (2 r_f \Eps_N^f + \sin 2\theta_{12} \epsilon_{12}^{-1})^2}},
\end{equation}
where $A \equiv -\cos 2\theta_{12} \epsilon_{12}^{-1} + c_{13}^2 -
2r_f \Eps_D^f$. The absolute minimum is achieved when the second term
in denominator of~\eqref{eq:asymptot} is zero, \textit{i.e.}
\begin{equation}
  \label{eq:gen}
  \Eps_N^f = - \frac{\sin 2 \theta_{12}}{2 r_f \epsilon_{12}}.
\end{equation}
This corresponds to zero the off-diagonal elements of
Hamiltonian~\eqref{eq:nsieff}.  For $E = 3~\text{MeV}$ we obtain
$\Eps_N^f = -0.3$.

In Fig.~\ref{fig:newphys} we plot the $P_{ee}$ survival probability
for non-standard interactions with up-quarks and down-quarks. As can
be seen, the presence of NSI allows to considerably flatten the
spectrum above $3~\text{MeV}$, in analogy with the similar effect
produced by light sterile neutrinos. Moreover, NSI can also generate
large day-night asymmetries: $(4 - 5)\%$ at $10~\text{MeV}$ for both
cases, $f = u$ and $f = d$.  The flattening and larger asymmetry
remove the tension with KamLAND data.

The NSI and sterile neutrino cases can be distinguished by the slower
increase of the NSI probability as the energy decreases. The sharp
increase for the sterile case is related to small $\nu_e - \nu_s$
mixing (narrow resonance). The two possibilities can be distinguished,
\textit{e.g.}, by precise measurements of the $pep$ neutrino flux and
well as the day-night asymmetry at high energies.

The NSI provide a very good fit to solar neutrino data, even in the
limit of $\Dmq_{21} = 0$. This is mainly due to the lack of
experimental data below $5~\text{MeV}$, where the transition between
the MSW and the vacuum-dominated regime takes place.  It is therefore
not possible to obtain a precise determination of \emph{both} vacuum
oscillation \emph{and} non-standard interaction parameters using only
solar data.
On the other hand, KamLAND measurement of $\Dmq_{21}$, being only
marginally affected by matter effects, is rather stable under the
presence of NSI. A combined fit of both solar and KamLAND data is
therefore able to constraint both sets of parameters with good
accuracy. In Fig.~\ref{fig:nonstand} we show the results of such a
combined fit for NSI with $f=u$ (left) and $f=d$ (right), limited for
simplicity to the case of real $\Eps_N^f$. The presence in both cases
of two disconnected regions is related to the ambiguity in the
determination of the sign of $\Dmq_{21}$ discussed above. Namely, the
left-side regions include the standard solution $\Eps_D^f = \Eps_N^f =
0$ $\Dmq_{21} > 0$, the right-side regions (slightly disfavored for
$f=d$) correspond to $\Dmq_{21} < 0$.

In Fig.~\ref{fig:nonstand} the best fit points (stars) for
solar+KamLAND are $\Eps_D^d = -0.12$ and $\Eps_N^d = -0.16$ for NSI
with down-type quarks, and $\Eps_D^u = -0.22$ and $\Eps_N^u = -0.30$
for NSI with up-type quarks.  These values are somewhat in tension
with the atmospheric and LBL experiments bounds, and probably hard to
accommodate within BSM models, although some possibility has been
discussed~\cite{Farzan:2015doa}. As shown in
Ref.~\cite{Gonzalez-Garcia:2013usa}, atmospheric and long-baseline
data are insensitive to $\Eps_D^f$ but more restrictive for
$\Eps_N^f$, and can therefore provide complementary information.

In addition to propagation effects, the presence of NSI can also
affect the neutrino cross-sections relevant for neutrino detection.
For example, in Ref.~\cite{Agarwalla:2012wf} stringent bounds on NSI
with electrons were derived from the Borexino data by studying the
elastic scattering in the \Nuc{7}{Be} energy window.


\subsection{Large magnetic moments}

Electromagnetic interactions of neutrinos are due to neutrino dipole
moments~(see review~\cite{Giunti:2014ixa}).  The Dirac neutrino in the
Standard Model has a very tiny magnetic moment, $\mu_\nu = 3\times
10^{-19} \, (m_\nu / \text{eV}) \, \mu_B$ Ref.~\cite{Fujikawa:1980yx}.
Experimental evidence for a larger value of $\mu_\nu$ would therefore
testify for the presence of some new physics physics.

Following the formalism of Ref.~\cite{Grimus:2000tq}, we describe the
interaction of Dirac or Majorana neutrinos with the electromagnetic
field in terms of an effective Hamiltonian in the flavor basis:
\begin{equation}
  \mathcal{H}_\text{em} =
  \begin{cases}
    \hphantom{-}\dfrac{1}{2} \bar\nu_R \lambda \sigma^{\alpha\beta} \nu_L
    F_{\alpha\beta} + \text{h.c.} &\text{(Dirac),}
    \\[3mm]
    -\dfrac{1}{4} \bar\nu_L^T C^{-1} \lambda \sigma^{\alpha\beta} \nu_L
    F_{\alpha\beta} + \text{h.c.} &\text{(Major.),}
  \end{cases}
\end{equation}
where $C$ is the charge-conjugation operator and $\nu_{L(R)}^T =
(\nu_e, \nu_\mu, \nu_\tau)_{L(R)}$ is the vector of left-handed
(right-handed) flavor states. The matrix $\lambda$ can be decomposed
into the sum of two Hermitian matrices:
\begin{equation}
  \lambda = \mu - i d \,,
  \quad
  \mu = (\lambda + \lambda^\dagger) / 2 \,,
  \quad
  d = i (\lambda - \lambda^\dagger) / 2 \,.
\end{equation}
Here $\mu$ describes the neutrino magnetic moments, while $d$ the
electric dipole moments. The off-diagonal elements of these matrices
link together states of opposite helicity and different flavors, and
thus correspond to the \emph{transition
  moments}~\cite{Schechter:1981hw}. For the \emph{Majorana} neutrinos
the CPT conservation implies that $\mu$ and $d$ are anti-symmetric
imaginary matrices, so that their diagonal elements vanish and only
transition moments are possible.

Bounds on the elements of the matrix $\lambda$ can be presented in
terms of the collective quantity $|\Lambda| =
\sqrt{\mathop{\mathrm{Tr}} (\lambda^\dagger \lambda) / 2}$.
For what concerns solar neutrinos, two effects of neutrino
electromagnetic properties have been considered: neutrino spin-flavor
precession and additional contribution to the neutrino-electron
scattering cross-section.


\subsubsection{Spin-flavor precession}

The evolution of neutrino states under the combined influence of
matter effects and strong magnetic fields, can be described by
equation~\cite{Lim:1987tk, Akhmedov:1988uk, Minakata:1988gm,
  Grimus:2000tq}:
\begin{equation}
  \label{eq:magnetic}
  i \frac{d}{dz}
  \begin{pmatrix}
    \varphi_- \\[2mm] \varphi_+
  \end{pmatrix}
  =
  \begin{pmatrix}
    \dfrac{M^\dagger M}{2E} + V_L & -B_+ \lambda^\dagger \\
    -B_- \lambda & \dfrac{M M^\dagger}{2E} + V_R
  \end{pmatrix}
  \begin{pmatrix}
    \varphi_- \\[2mm] \varphi_+
  \end{pmatrix},
\end{equation}
where $\varphi_+$ and $\varphi_-$ denote the vectors of neutrino
flavor states corresponding to positive and negative helicities,
respectively. We have assumed that neutrino propagate along the $z$
direction, so that $B_\pm = B_x \pm i B_y$ are the components of the
magnetic field perpendicular to the neutrino trajectory. Here $V_L =
\diag(V_e + V_n, V_n, V_n)$ is the standard matter potential for
neutrino states; for Dirac neutrinos we have $V_R = 0$, while for
Majorana neutrinos we have $V_R = -V_L$.

The joint evolution of flavor and spin states induced by the existence
of magnetic transition moments leads to the phenomenon of spin-flavor
precession~\cite{Lim:1987tk, Akhmedov:1988uk, Minakata:1988gm}.  The
interplay of such mechanism and standard matter effects could be the
source of the observed deficit of solar neutrinos, as long as
$10^{-9}~\eVq \lesssim \Dmq_{21} \lesssim
10^{-7}~\eVq$~\cite{Miranda:2001hv, Akhmedov:2002mf}.  However, the
evidence for much larger $\Dmq_{21}$ provided by KamLAND, which is
practically insensitive to small neutrino magnetic moment, ruled out
this mechanism. Conversely, we can now use the precise determination
of the oscillation parameters to set bounds on neutrino
electromagnetic properties, by requiring that the accurate description
of solar neutrino data is not spoiled by spin-flavor precession
effects, and by the fact that no antineutrino coming from the sun is
detected. For example, in Ref.~\cite{Miranda:2004nz} the bound
$|\Lambda| \lesssim \text{few}\times 10^{-12}\,\mu_B$ for $\theta_{13}
= 0$ was derived, under the assumption that neutrinos are Majorana
particles and that turbulent random magnetic fields exists in the Sun.

As can be seen in Eq.~\eqref{eq:magnetic}, in the presence of
non-vanishing magnetic moment the evolution of the neutrino system
mixes together positive and negative helicities, so that a conversion
between them becomes possible. If neutrinos are Majorana particles
this implies that neutrinos can convert into anti-neutrinos, which may
result in the observation of a flux of anti-neutrinos coming from the
Sun. Searches for $\bar{\nu}_e$ signal have been performed by both
Borexino~\cite{Bellini:2010gn} and KamLAND~\cite{Eguchi:2003gg}, so
far with negative results.

As shown in Ref.~\cite{Miranda:2001hv}, the large value of
$\Dmq_{21}$, as measured by KamLAND, implies that a neutrino magnetic
moment below $10^{-11}~\eVq$ has practically no effect on the
evolution of solar neutrinos, given a characteristic solar magnetic
field of the order of 80~kG.


\subsubsection{Neutrino-electron cross-section}

In the presence of magnetic moments extra term arises in the elastic
neutrino-electron cross-section due to photon exchange:
\begin{equation}
  \frac{d\sigma_\text{em}}{dT}
  = \frac{\alpha^2 \pi}{m_e^2 \mu_B^2}
  \left( \frac{1}{T} - \frac{1}{E} \right)
  \Big(
  \lVert \lambda \varphi_-^\text{det} \rVert^2
  + \lVert \lambda^\dagger \varphi_+^\text{det} \rVert^2
  \Big),
\end{equation}
where $E$ is the neutrino energy, $T$ is the kinetic energy of the
recoil electron, and the 3-vectors $\varphi_-^\text{det}$ and
$\varphi_+^\text{det}$ denote the neutrino flavor amplitudes at the
detector for negative and positive helicities.

In the case small precession effects in the Sun, the helicity of solar
neutrinos is conserved, so that $\varphi_+^\text{det} = 0$ and
$\varphi_-^\text{det}$ can be calculated using the formalism
introduced in Sec.~\ref{sec:msw}. Following this formalism, in
Ref.~\cite{Grimus:2002vb} a bound on Majorana transition moments was
derived, $|\Lambda| \lesssim 6.3\times 10^{-10}\,\mu_B$ from the
analysis of solar neutrino data alone, and $|\Lambda| \lesssim
2.0\times 10^{-10}\,\mu_B$ in combination with reactor antineutrino
data. Such a bound can be improved by almost an order of magnitude
after the inclusion of 3 years of Borexino data.


\subsection{Neutrino decay}

The existence of neutrino masses and flavour mixing implies that the
heavier neutrino states decay into lighter ones, and are therefore
unstable~\cite{Bahcall:1972my}. In the Standard Model, the neutrino
lifetimes are much longer than the age of the Universe, hence well
beyond the reach of present experiments.  Observation of neutrino
decay would therefore be a signal of New Physics.

In vacuum, the survival probability of an unstable state $\nu_i$ is
described by an exponential factor $e^{-d_i L/E}$, where $E$ is the
neutrino energy, $L$ is the traveled distance, and $d_i = m_i /
\tau_i$ is the ratio of the neutrino mass and lifetime. Applying this
to solar neutrinos and neglecting decay inside the Sun or across the
Earth, we obtain
\begin{equation}
  P_{e\alpha} = \sum_i P_{ei}^S P_{i\alpha}^E e^{-d_i L_{SE}/E},
\end{equation}
where $P_{ei}^S$ and $P_{i\alpha}^E$ are the $\nu_e \to \nu_i$ and
$\nu_i \to \nu_\alpha$ probabilities in the Sun and the Earth,
respectively (see Sec.~\ref{sec:evol-sun}
and~\ref{sec:evol-earth}). If the decay daughter particles include
lighter active neutrinos, they should be accounted for in the
calculation of the event rates. Here we follow the approach of
Refs.~\cite{Berryman:2014qha, Picoreti:2015ika} and we ignore this
possibility, thus assuming that decaying neutrinos simply
``disappear''.

Due to the smallness of $\theta_{13}$ the impact of a nonzero $d_3$ on
solar neutrino data is very small~\cite{Berryman:2014qha}. Indeed,
from a global fit of the solar neutrino data we find that for
$\sin^2\theta_{13} = 0.022$ the values $d_3 \gg 10^{-10}~\eVq$
(complete $\nu_3$ decay) are disfavored with respect to $d_3 \ll
10^{-13}~\eVq$ (stable $\nu_3$ state) by $\Delta\chi^2=0.55$
only. Hence, no bound can be set on $d_3$ from the present data.  On
the other hand, from the same analysis we find $d_1 = m_1/\tau_1 <
1.3\times 10^{-13}~\eVq$ and $d_2 = m_2/\tau_2 < 1.2\times
10^{-12}~\eVq$ at the $3\sigma$ level, in a good agreement with the
results of Refs.~\cite{Berryman:2014qha, Picoreti:2015ika}. The global
best fit point occurs for $d_i = 0$, and the determination of the
$\theta_{12}$ range is practically unaffected by the enlargement of
the parameter space.


\subsection{Violation of fundamental symmetries}

Violation of fundamental symmetries (VFS) at the Planck mass scale is
expected in theories attempting to unify gravity with quantum physics.
Neutrinos, whose masses are generated by some physics close to the
GUT/Planck scales, could be most sensitive to this violation.  New
effects in neutrino propagation may arise due to violations of the
equivalence principle~\cite{Gasperini:1988zf}, neutrino couplings to
space-time torsion fields~\cite{DeSabbata:1981ek}, violation of
Lorentz invariance~\cite{Coleman:1997xq} and of CPT
symmetry~\cite{Colladay:1996iz}.  The impact of VFS on neutrino
propagation can be accounted by the inclusion of extra terms in the
Hamiltonian:
\begin{equation}
  \label{eq:npv}
  H_\textsc{np}^{(2)}
  = \frac{1}{2} \sigma^\pm E^n
  \begin{pmatrix}
    -\varphi_D & ~\varphi_N \\
    \hphantom{+} \varphi_N^* & ~\varphi_D
  \end{pmatrix},
\end{equation}
where $n$ is the energy power scaling, $\varphi_D$ and $\varphi_N$
parametrize the strength of the VFS effects, and $\sigma^\pm$ accounts
for a possible relative sign between neutrinos and antineutrinos. We
have ($n=1$, $\sigma^+ = \sigma^-$) in the case of violation of the
equivalence principle or violation of Lorentz invariance, ($n=0$,
$\sigma^+ = \sigma^-$) for neutrino couplings to space-time torsion
fields, and ($n=0$, $\sigma^+ = -\sigma^-$) for violation of CPT
symmetry.  $H_\textsc{np}$ does not involve matter or magnetic fields
and therefore relevant also for propagation in \emph{vacuum}.

The oscillation probabilities in the presence of VFS can be derived
from the standard ones by just replacing the vacuum mixing and
mass-squared splittings with effective quantities defined by
diagonalizing the vacuum Hamiltonian.
VFS can manifest itself through (A) deviation energy dependence of the
oscillation length from $E^{-1}$, and (B) non-trivial dependence of
the vacuum mixing parameters on the neutrino energy. So spectral
information is necessary for searches of VFS.

Concerning solar neutrinos, a bound on the overall strength of VFS,
$||\varphi|| = \sqrt{|\varphi_D|^2 + |\varphi_N|^2}$, can be estimated
by requiring that the VFS term of Eq.~\eqref{eq:npv} is not larger
than the standard one, $||\varphi|| E^n \lesssim \Dmq_{21} / E$. Using
$\Dmq_{21} = \mathcal{O}(10^{-5}~\eVq)$ and assuming a typical energy
$E \approx 10~\text{MeV}$ (the scale at which spectral information is
available from SK and SNO) we obtain $||\varphi|| \lesssim
10^{-12}~\text{eV}$ for $n=0$ and $||\varphi|| \lesssim 10^{-19}$ for
$n=1$. Such estimations are consistent with results of numerical
calculations.


\subsection{Other new physics models}

Practically any extension of the Standard Model, which leads to
non-standard neutrino properties, produces observable effects in the
neutrino oscillation pattern. The number of extensions which could be
probed by solar neutrinos is therefore huge, and here we briefly
mention few cases.
\begin{enumerate}
\item Mass Varying neutrinos were proposed in
  Ref.~\cite{Fardon:2003eh} to provide a theoretical framework for the
  otherwise unexplained closeness of values of the dark energy and
  dark matter densities today, even though their ratio varies in time
  as the third power of the cosmic scale factor. The model proposes
  that the dark energy and neutrino densities track each other, and
  that the neutrino mass is not a constant but rather a dynamical
  quantity arising from the minimization of an effective potential
  depending solely on the neutrino density itself.

  The phenomenological implications of this model for solar neutrinos
  were discussed in Ref.~\cite{Cirelli:2005sg}, where it was shown
  that the quality of the data fit is worse than in the standard
  case. A modification of the original model in which neutrino masses
  depend also on the density of visible matter was proposed
  in~\cite{Kaplan:2004dq}, and it was shown in~\cite{Barger:2005mn}
  that such a model is compatible with solar neutrino data.

\item The possible existence of other long-range forces beyond the
  electromagnetic and gravitational ones was first considered in
  Ref.~\cite{Lee:1955vk}. The phenomenological implications for solar
  neutrinos of a new leptonic force of this kind was discussed in
  Ref.~\cite{GonzalezGarcia:2006vp}. It was shown that such scenarios
  did not provide significant improvement of quality of the fit with
  respect to the standard LMA solution, and bounds on the strength and
  range of the new force were derived.

\item Non-standard decoherence effects are usually expected to be a
  possible manifestation of quantum gravity, for example in the
  presence of a ``foamy'' space-time fabric. The phenomenological
  implications of this mechanism for oscillating systems were first
  discussed in Ref.~\cite{Ellis:1983jz}. A concrete analysis applied
  to solar neutrinos was presented in~\cite{Fogli:2007tx}, where it
  was shown that the existence of non-standard sources of decoherence
  would induce extra smearing in the neutrino oscillation pattern and
  could therefore be detected experimentally. A fit to the available
  data showed no hint for such an effect, and stringent bounds on the
  new physics parameters were therefore derived.

\item The impact of extra dimensions on neutrino physics was first
  considered in~\cite{ArkaniHamed:1998vp, Dvali:1999cn}. These models
  share with sterile neutrino models the idea that extra fermionic
  singlets may exist, but allow them to propagate into a
  higher-dimensional spacetime whereas active neutrinos are confined
  to a (3+1) brane. The applications for solar neutrinos were
  discussed in~\cite{Dvali:1999cn, Caldwell:2001dj, Caldwell:2000zn,
    Lukas:2000wn}. The most distinctive property of such models is the
  presence of an infinite Kaluza-Klein tower of new neutrino
  eigenstates, which participate in the oscillation process and may
  therefore produce new MSW resonances.
\end{enumerate}


\section{Conclusion and Outlook}

Solar neutrino studies triggered vast developments in neutrino
physics.  The solar neutrino problem has been uncovered, and
eventually resolved in terms of the neutrino flavor conversion.
Theory of neutrino propagation in different media has been elaborated
which included the MSW effect (adiabatic flavor conversion), resonance
enhancement of oscillations, neutrino spin precession, resonance
spin-flavor precession, conversion in the presence of non-standard
interactions, \textit{etc}.  Effects of propagation in different
density profiles have been explored; among them the non-adiabatic
conversion, parametric effects, in particular, parametric enhancement
of neutrino oscillations, propagation in stochastic media, multi-layer
media, \textit{etc}.

Solar neutrinos played crucial role in establishing the standard
$3\nu$ paradigm with mixing of 3 flavors.
They provided determination of the 1-2 mixing and mass splitting,
fixed the sign of $\Dmq_{21}$, \textit{i.e.}, determined the 1-2 mass
hierarchy. They give independent measurements of $\theta_{13}$.

Physics beyond three neutrinos can show up in solar neutrinos as
sub-leading effects. The bound have been obtained on NSI, magnetic
moments of neutrinos, parameters of hypothetical sterile neutrinos,
neutrino decay, Lorentz violation and CPT violation parameters.  Some
of these bounds are the best or competitive with bounds obtained from
non-solar neutrino experiments.  The Sun here appears as a source of
neutrinos for various searches.

LMA MSW solution gives consistent description of all the data.  The
largest pulls are related to flat suppression of the flux at low
energies instead of upturn and slightly larger DN asymmetry.  Mixing
parameters extracted from the solar neutrino data are in agreement
with KamLAND results and consistent with the 1-3 mixing measurements
at reactors.  Although the $\Dmq_{21}$ determined from solar neutrinos
is about $2\sigma$ smaller than from KamLAND. The difference is
related to the absence of upturn and large D-N asymmetry.  This can be
just statistical fluctuation or indication of some new physics like
contribution from NSI to the matter potential or existence of very
light sterile neutrinos.

Measurements of $\Dmq_{21}$ and $\theta_{12}$ have crucial
implications for the fundamental theory.  They created new theoretical
puzzle --~large mixing with deviation from maximal mixing by about
Cabibbo angle.

Two experimental results obtained in 2014: direct measurements of the
$pp$-neutrino flux by Borexino and establishing at about $3\sigma$
level the Day-Night effect have accomplished the first phase of
studies of solar neutrinos. The next phase is precision (at sub $\%$
level) measurements of neutrino signals.  New opportunities are
related to SNO+, JUNO, HK:
\begin{enumerate}[label=\alph*)]
\item Accurate measurements of $pp$-, $pep$- and \Nuc{7}{Be}-neutrino
  fluxes will substantially contribute to global fits of the solar
 neutrino  data and to further checks of the LMA solution.

\item Detailed studies of the Earth matter effects will be possible
  using the Hyper-Kamiokande detector with possible applications to
  the Earth tomography.

\item In combination with other measurements solar neutrino studies
  provide sensitive way to test theory of neutrino oscillations and
  flavor conversion in matter.

\item Searches for sub-leading effects will allow to put more
  stringent bounds on non-standard neutrino properties.  In new phase
  of the field many small-size effects will be accessible and can not
  be neglected as before.
\end{enumerate}

Detailed knowledge of solar neutrinos is needed for various low
background experiments, in particular, for searches of the Dark matter
and neutrinoless double beta decay.  In future the solar neutrino
fluxes will be accessible to the Dark Matter
detectors~\cite{Billard:2014yka}.


\section*{Acknowledgements}

A.S.\ would like to thank the Instituto de Fisica Teorica (IFT
UAM-CSIC) in Madrid, where this work developed, for its support via
the Severo Ochoa Distinguished Visiting Professor position.
This work is supported by Spanish MINECO grants FPA2012-31880 and
FPA2012-34694, by the Severo Ochoa program SEV-2012-0249 and
consolider-ingenio 2010 grant CSD-2008-0037, and by EU grant FP7 ITN
INVISIBLES (Marie Curie Actions PITN-GA-2011-289442).


\end{document}